\documentclass[conference]{IEEEtran}
\pdfoutput=1

\usepackage{cite}
\usepackage{amsmath,amssymb,amsfonts}
\usepackage{algorithmic}
\usepackage{graphicx}
\usepackage{textcomp}
\usepackage{xcolor}
\usepackage{fancyhdr}
\usepackage{multirow}
\usepackage{ifpdf}
\usepackage[hyphens]{url}

\IEEEoverridecommandlockouts

\def\BibTeX{{\rm B\kern-.05em{\sc i\kern-.025em b}\kern-.08em
    T\kern-.1667em\lower.7ex\hbox{E}\kern-.125emX}}

\fancypagestyle{firstpage}{
  \fancyhf{}

  \fancyfoot[C]{\thepage}
}

\newif\ifrev
\revtrue
\ifrev
\newcommand{\fixme}[1]{\textcolor{red}{(ToDo: #1)}}

\else
    \newcommand{\fixme}[1]{}

\fi

\newcommand{\DesName}{{\textit{RecNMP}~}}
\newcommand{\DesNameni}{{RecNMP~}}

\title{RecNMP: Accelerating Personalized Recommendation with Near-Memory Processing
\vspace{-0.0em}
}

\author{\normalsize Liu Ke$^\ast$\thanks{$^\ast$Washington University in St. Louis, work done while at Facebook.},
Udit Gupta, Carole-Jean Wu, Benjamin Youngjae Cho, Mark Hempstead, Brandon Reagen, Xuan Zhang$^\ast$\\
\\
\normalsize David Brooks, Vikas Chandra, Utku Diril, Amin Firoozshahian, Kim Hazelwood, Bill Jia, Hsien-Hsin S. Lee\\
\normalsize Meng Li, Bert Maher, Dheevatsa Mudigere, Maxim Naumov, Martin Schatz, Mikhail Smelyanskiy, Xiaodong Wang\\
\\
\normalsize Facebook, 1 Hacker Way, Menlo Park, CA 94025 \\
\normalsize liuke@fb.com, xuan.zhang@wustl.edu
}

\begin{document}
\maketitle
\thispagestyle{firstpage}
\pagestyle{plain}


\begin{abstract}

Personalized recommendation systems leverage deep learning models and account for the majority of data center AI cycles.
Their performance is dominated by memory-bound sparse embedding operations with unique irregular memory access patterns that pose a fundamental challenge to accelerate.
This paper proposes a lightweight, commodity DRAM compliant,
near-memory processing solution to accelerate personalized recommendation inference.
The in-depth characterization of production-grade
recommendation models shows that embedding operations with high model-, operator- and data-level parallelism lead to memory bandwidth saturation, limiting recommendation inference performance.
We propose \textit{RecNMP} which provides a scalable solution to improve system throughput, supporting a broad range of sparse
embedding models. \textit{RecNMP} is specifically tailored to production environments with heavy co-location of operators on a single server.
Several hardware/software co-optimization techniques such as memory-side caching, table-aware packet scheduling, and hot entry profiling are studied, resulting in up to 9.8$\times$ memory latency speedup over a highly-optimized baseline.
Overall, \textit{RecNMP} offers 4.2$\times$ throughput improvement and 45.8\% memory energy savings.

\end{abstract}

\section{Introduction}
\label{sec:intro}
\vspace{-0.1cm}


Personalized recommendation is a fundamental building block of 
many internet services used by search engines, social networks, 
online retail, and content streaming~\cite{GCP,Covington:2016,Walmart_AI,Amazon_Personalize}. 
Today's personalized recommendation systems leverage 
deep learning to maximize accuracy and deliver the best user experience~\cite{hazelwood2018applied,Zhou:2018,Guo:2018,Cheng:2016,DLRM}. 
The underlying deep learning models 
now consume the majority of the datacenter cycles spent on AI~\cite{arxiv-gupta-19,sigarch-blog}.
For example, recent analysis reveals that the top recommendation models collectively contribute to more than 72\% of all AI inference cycles across Facebook's production datacenters~\cite{sigarch-blog}.

Despite the large computational demand and production impact, 
relatively little research has been conducted to optimize deep learning (DL)-based recommendation.
Most research efforts within the architecture community have
focused on accelerating the compute-intensive, highly-regular computational patterns found in
fully-connected (FC), convolution (CNN), and recurrent (RNN) 
neural networks~\cite{cnvlutin,ISAAC,EIE,Minerva,Eyeriss,NeuroCube,Cambricon,TPU_short,ScaleDeep,SCNN,MacCNNEfficiency,Scalpel,OptLPSGD,DNNmemristorreliable,GANaccelerator,CompressingDMA,InSituAI,Ganax,snapea,UCNN,OutlierDNN,PredictionCNN,Bitfusion,Gist,DarkPruning,InputSim}.
Unlike CNNs and RNNs, recommendation models exhibit low compute-intensity and little to no regularity.
Existing acceleration techniques either do not apply
or offer small improvements at best, as they tend to exploit regular reusable dataflow patterns and assume high spatial locality which are not the main performance bottleneck in recommendation models~\cite{arxiv-gupta-19}.
Given the volume of personalized inferences and their rapid growth rate occurring in the data center,
an analogous effort to improve performance of these models would have substantial impact.

\begin{figure}[t!]
  \centering
  \includegraphics[width=\columnwidth]{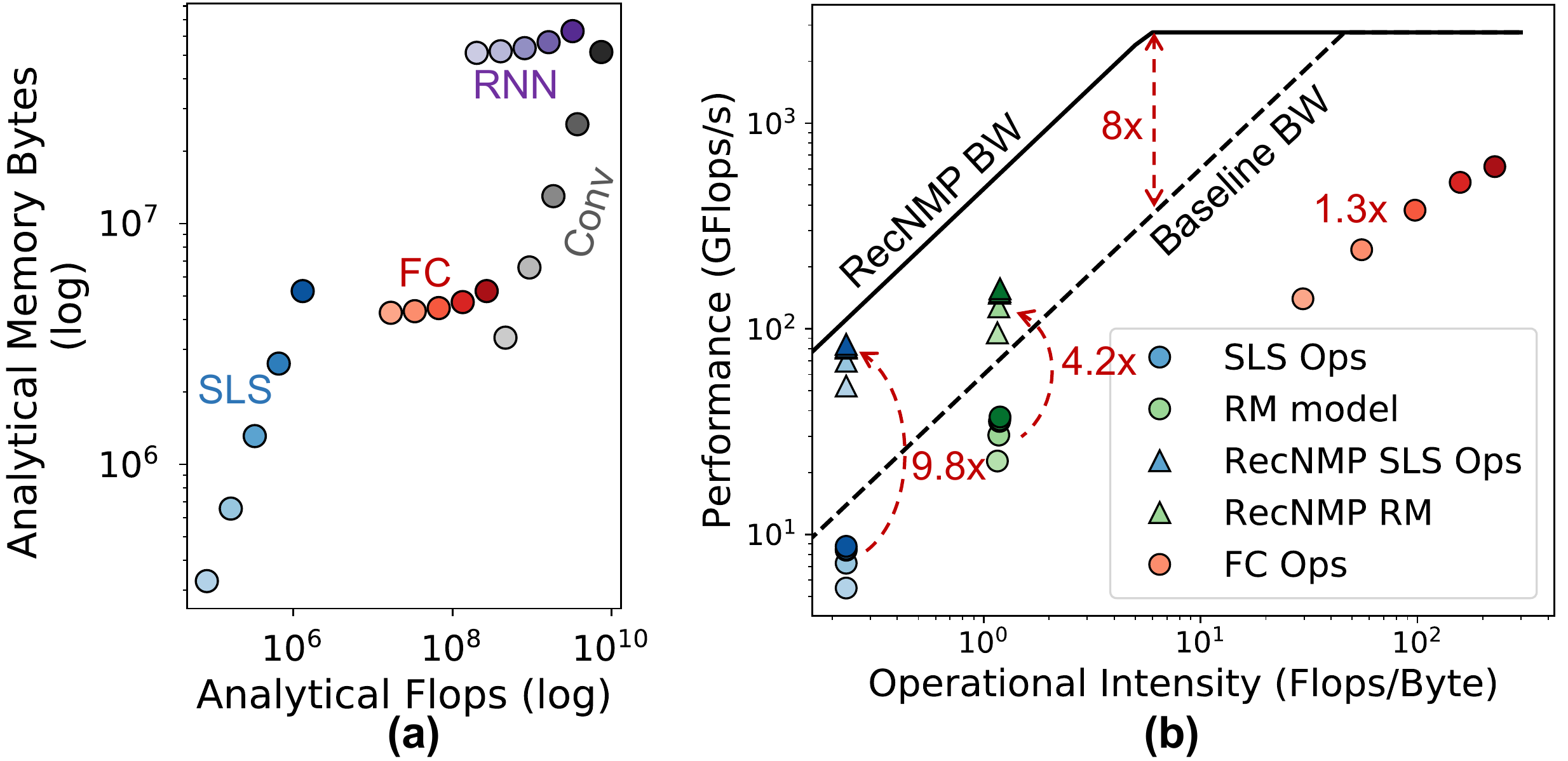}
  \vspace{-0.7cm}
  \caption{(a) Compute and memory footprint of common deep learning operators, sweeping batch size; 
            (b) Roofline lifting effect and the operator-level (FC, SLS) and end-to-end model (RM) speedup enabled by \textit{RecNMP}.}
  \label{fig:motiv}
  \vspace{-0.6cm}
\end{figure}

%
To suggest personalized contents to individual users, recommendation models are generally structured to take advantage of both continuous (dense) and categorical (sparse) features.
The latter are captured by large embedding tables with sparse lookup and pooling operations.
These embedding operations dominate the run-time of 
recommendation models and are markedly distinct from other layer types.

A quantitative comparison of the raw compute and memory access requirements is
shown in Figure~\ref{fig:motiv}(a).
Sparse embedding operations, represented by SparseLengthsSum (SLS), consist of a small sparse lookup into a large embedding table
followed by a reduction of the embedding entries (i.e., pooling).
They present two unique challenges:
First, while the sparse lookup working set is comparatively small (MBs), the irregular nature of the table indices exhibits poor predictability, rendering typical prefetching and dataflow optimization techniques ineffective.
Second, the embedding tables are on the order of tens to hundreds of GBs,
overwhelming on-chip memory resources.
Furthermore, the circular points in Figure~\ref{fig:motiv}(b) show the operational
intensity of SLS is orders of magnitude less than FC layers.
Low intensity limits the potential of custom hardware 
including the specialized datapaths and on-chip memories used in CNN/RNN accelerators.
The result is a fundamental memory bottleneck that cannot be overcome
with standard 
caching (e.g., tiling~\cite{tile}),
algorithmic (e.g., input batching),
or hardware acceleration techniques.

This paper proposes \textit{RecNMP}---a near-memory processing solution
to accelerate the embedding operations for DL-based recommendation.
\DesName is a lightweight DIMM-based system built on top of existing standard DRAM technology. 
We focus on DIMM-based near-memory processing~\cite{nda-kim,chameleon,tensorDIMM}
instead of resorting to specialized 2.5D/3D integration processes (e.g. HBM)~\cite{graphpim,pim-enabled,NeuroCube}. The DIMM form factor with commodity DDR4 
devices can support the 100GB+ capacities necessary for production-scale recommendation models with low cost. 
By eliminating the off-chip memory bottleneck and exposing higher internal  bandwidth
we find that \DesName provides significant opportunity to improve performance and efficiency by lifting the roofline by 8$\times$ for the bandwidth-constrained region 
(Figure~\ref{fig:motiv}(b)), enabling optimization opportunity not feasible with existing 
systems.

We have performed a detailed characterization of recommendation models using open-source, production-scale DLRM benchmark~\cite{DLRM, arxiv-gupta-19} as a case study.
This analysis quantifies the potential benefits of near-memory processing in accelerating recommendation models and builds the intuition for co-designing the NMP hardware with the algorithmic properties of recommendation.
Specifically, it highlights the opportunity for the \DesName architecture in which
bandwidth-intensive embedding table operations are performed in the memory and compute-intensive
FC operators are performed on the CPU (or potentially on an accelerator).

The proposed \DesName design exploits DIMM- and rank-level parallelism in DRAM memory systems. \DesName performs local lookup and pooling functions near memory, supporting a range of sparse embedding inference operators, which produces the general Gather-Reduce execution pattern.
In contrast to a general-purpose NMP architecture, we make a judicious design choice to implement selected lightweight functional units with small memory-side caches to limit the area overhead and power consumption. We combine this light-weight hardware with software optimizations including table-aware packet scheduling and hot entry profiling.
Compared to previous work whose performance evaluation is solely based on randomly-generated embedding accesses~\cite{tensorDIMM}, our characterization and experimental methodology is modeled after representative production configurations and is evaluated using real production embedding table traces.
Overall, \DesName leads to significant embedding access latency reduction ($9.8\times$) and improves end-to-end recommendation inference performance ($4.2\times$) as illustrated in Figure~\ref{fig:motiv}(b). 
Our work makes the following research contributions:
\begin{itemize}
 \setlength{\itemsep}{0pt}
 \setlength{\parskip}{0pt}






\item Our in-depth workload characterization shows that production recommendation models are constrained by memory bandwidth.
Our locality analysis using production embedding table traces reveals distinctive spatial and temporal reuse patterns and motivates a custom-designed NMP approach for recommendation acceleration.

\item We propose \textit{RecNMP}, a lightweight DDR4-compatible near-memory processing architecture. \DesName accelerates the execution of a broad class of recommendation models and exhibits 9.8$\times$ memory latency speedup and 45.9\% memory energy savings.
Overall, \DesName achieves 4.2$\times$ end-to-end throughput improvement.

\item We examine \textit{hardware-software co-optimization} techniques (memory-side caching, table-aware packet scheduling, and hot entry profiling) to enhance \DesName performance, and customized NMP instruction with $8\times$ DRAM command/address bandwidth expansion.

\item A \textit{production-aware evaluation framework} is developed to take into account common data-center practices and representative production configuration, such as model co-location and load balancing.

\end{itemize}

\section{Characterizing Deep Learning Personalized Recommendation Models}
\label{sec:char}

This section describes the general architecture of DL-based recommendation
models with prominent sparse embedding features and their performance bottlenecks. 
As a case study, we conduct a thorough characterization of the recently-released Deep Learning Recommendation Model (DLRM) benchmark~\cite{DLRM}. 
The characterization---latency breakdown, roofline analysis, bandwidth analysis, and memory locality---illustrates the unique memory requirements and access behavior of production-scale recommendation models and justifies the proposed near-memory accelerator architecture.


\vspace{-0.1cm}
\subsection{Overview of Personalized Recommendation Models} 
\label{sec:recModels}
\vspace{-0.1cm}
Personalized recommendation is the task of recommending content to users based on their preferences and previous interactions.
For instance, video ranking (e.g., Netflix, YouTube), a small number of videos, out of potentially millions, must be recommended to each user.
Thus, delivering accurate recommendations in a timely and efficient manner is important. 

\begin{figure}[t!]
 \centering
  \includegraphics[width=\columnwidth]{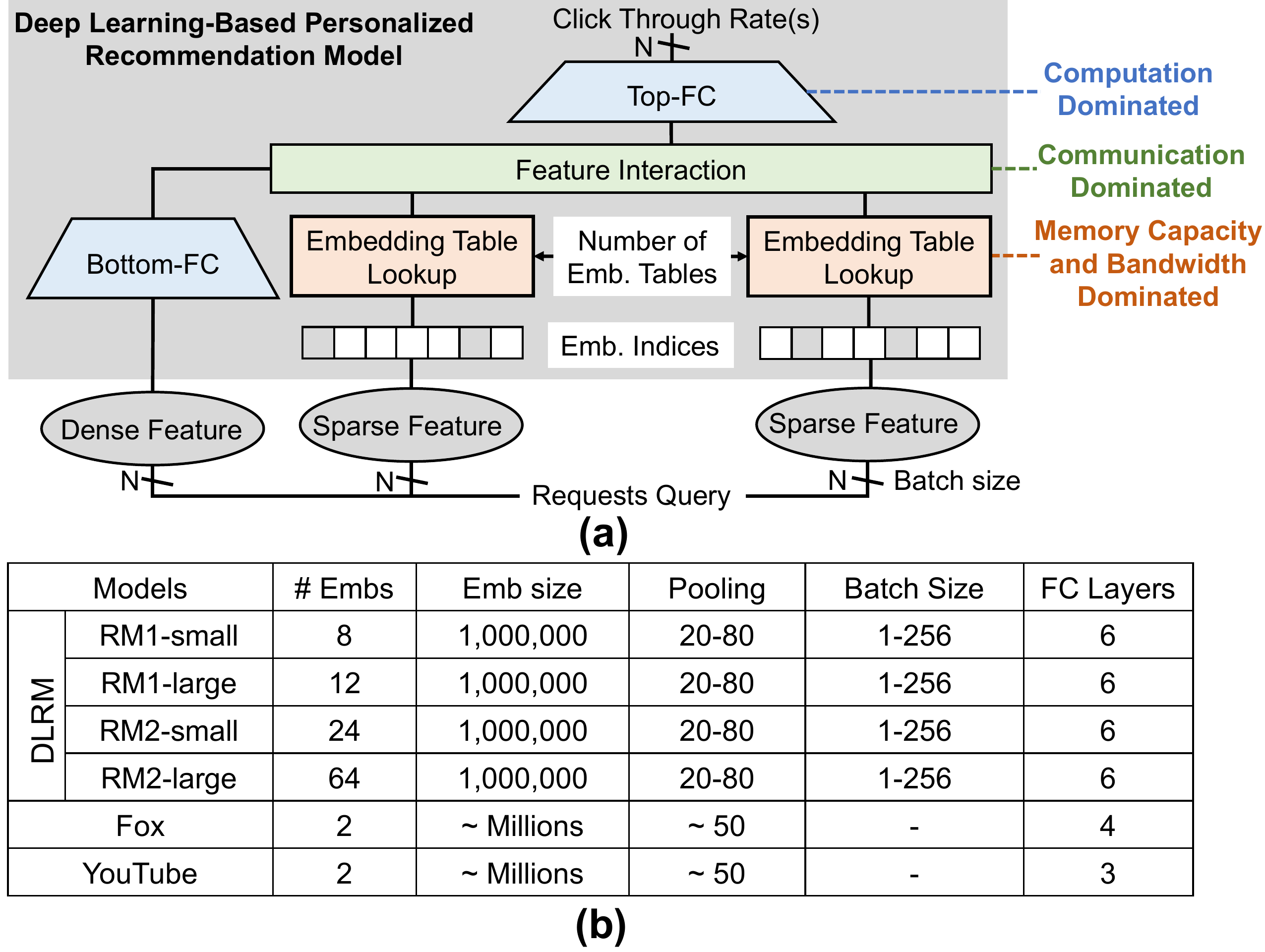}
  \vspace{-0.7cm}
  \caption{
  (a) Simplified model-architecture reflecting production-scale recommendation models;
  (b) Parameters of representative recommendation models.}
  \label{fig:sparsenn}
  \vspace{-0.6cm}
\end{figure}

Most modern recommendation models have an extremely large feature set to capture a range of user behavior and preferences. 
These features are typically separated out into dense and sparse features. 
While dense features (i.e., vectors, matrices) are processed by typical DNN layers (i.e., FC, CNN, RNN), sparse features are processed by indexing large embedding tables. 
A general model architecture of DL-based recommendation systems is captured in Figure~\ref{fig:sparsenn}. A few examples are listed with their specific model parameters~\cite{DLRM,fox,youtube} in Figure~\ref{fig:sparsenn}(b).
Similar mixture of dense and sparse features are broadly observable across many alternative recommendation models~\cite{alibabaRec,wideanddeep,mtwnd,DLRM,fox,youtube}.

Embedding table lookup and pooling operations provide an abstract representation of sparse features learned during training and are central to DL-based recommendation models. 
Embedding tables are organized as a set of potentially millions of vectors.
Generally, embedding table operations exhibit Gather-Reduce pattern; the specific element-wise reduction operation varies between models. 
For example, Caffe\cite{caffe2} comprises a family of embedding operations, prefixed by \textit{SparseLengths} (i.e., SparseLengthsWeightedSum8BitsRowwise), that perform a similar Gather-Reduce embedding operation with quantized, weighted summation. 
The SLS operator primitive is widely employed by other production-scale recommendation applications (e.g. YouTube~\cite{youtube} and Fox~\cite{fox}).
Our work aims to alleviate this performance bottleneck and improve system throughput by devising a novel NMP solution to offload the SLS-family embedding operations thus covering a broad class of recommendation systems.

\vspace{-0.1cm}
\subsection{A Case Study---Facebook's DLRM Benchmark}
\vspace{-0.1cm}

To demonstrate the advantages of near-memory processing for at-scale personalized recommendation models, we study Facebook's deep learning recommendation models (DLRMs)~\cite{DLRM}.
Dense features are initially processed by the BottomFC operators, while sparse input features are processed through the embedding table lookups.
The output of these operatiors are combined and processed by TopFC producing a prediction of click-through-rate of the user-item pair.

This paper focuses on performance acceleration strategies for four recommendation models representing two canonical classes of the models, RM1 and RM2~\cite{arxiv-gupta-19}.
These two model classes attribute to significant machine learning execution cycles at Facebook's production datacenter, RM1 over 30\% and RM2 over 25\%~\cite{sigarch-blog}. 
The parameters to configure them are shown in Figure~\ref{fig:sparsenn}(b).
The notable distinguishing factor across these configurations is the number of the embedding tables.
RM1 is a comparatively smaller model with few embedding tables;
RM2 has tens of embedding tables.

\begin{figure}[t]
  \centering
  \includegraphics[width=\columnwidth]{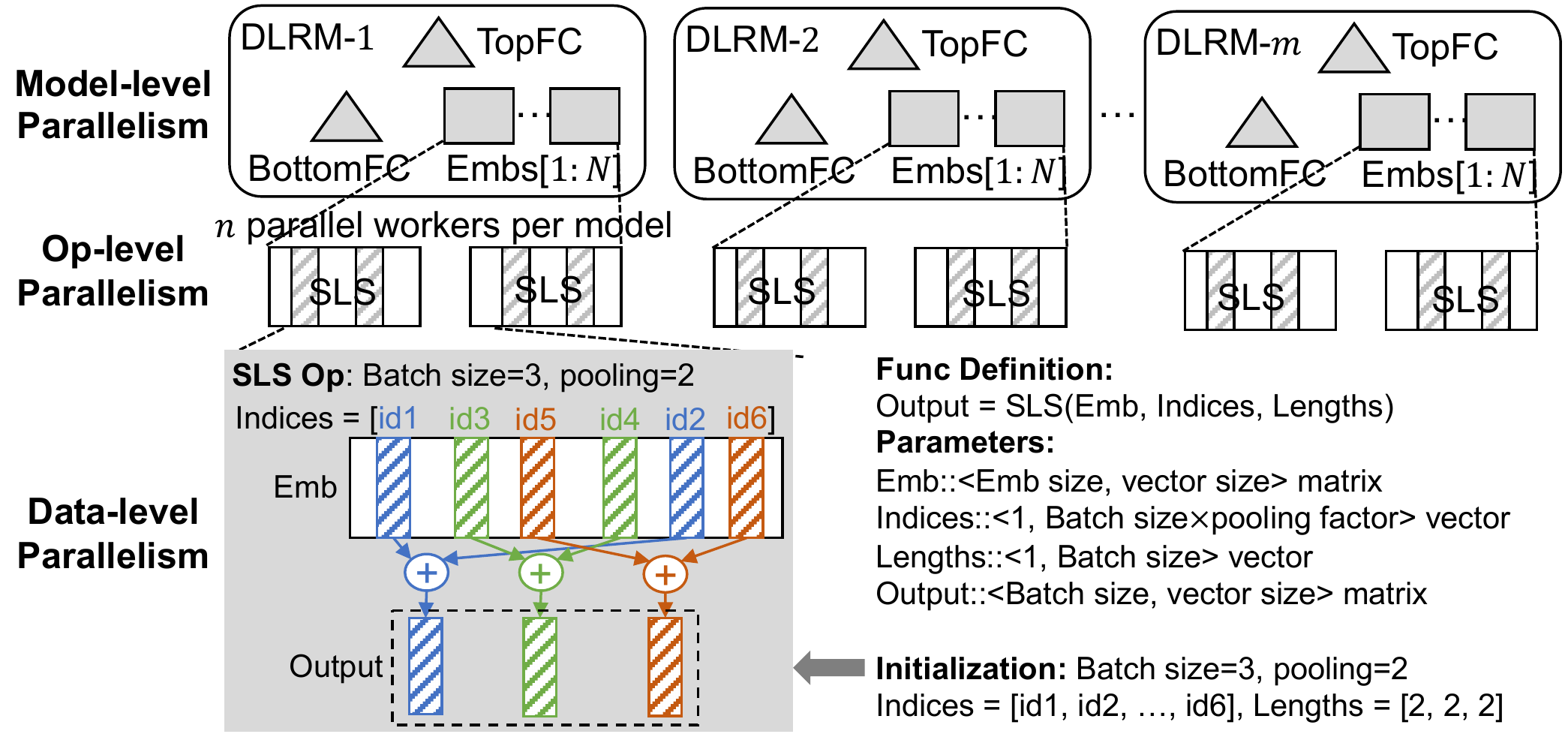}
  \vspace{-0.7cm}
  \caption{Model-, operator- and data-level parallelism in production system.}
  \label{fig:model_parallelism}
  \vspace{-0.3cm}
\end{figure}

In production environments, recommendation models employ three levels of parallelism, shown in Figure~\ref{fig:model_parallelism}, to achieve high throughput under strict latency constraints~\cite{arxiv-gupta-19}. 
Model-level parallelism grows by increasing the number of concurrent model inference  ($m$) on a single machine, operator-level parallelism adds parallel threads ($n$) per model and data-level parallelism is scaled by increasing batch size.
An SLS operator performs a batch of pooling operations; one pooling operation performs the summation for a set of vectors.
The inputs to SLS, for one batch of embedding lookups, include an indices vector containing sparse-IDs, and optionally a weight vector.


\vspace{-0.1cm}
\subsection{Operator Bottleneck Study}
\vspace{-0.1cm}

\begin{figure}[t!]
  \centering
  \includegraphics[width=\columnwidth]{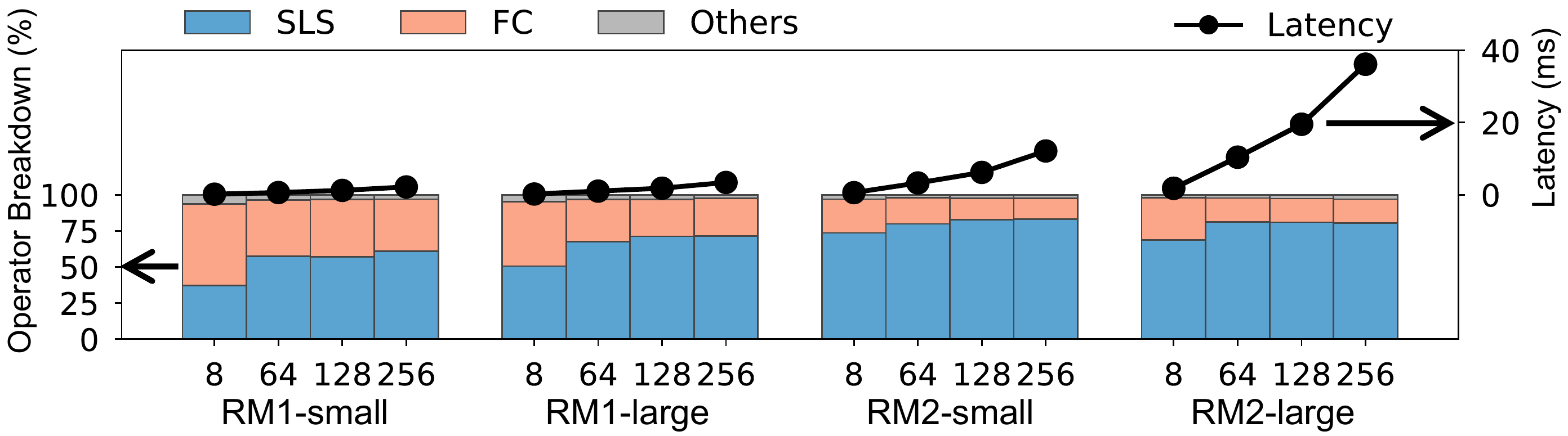}
  \vspace{-0.8cm}
  \caption{Inference latency and breakdown across models (RM1-small, RM1-large, RM2-small, RM2-large) with varying batch sizes (8, 64, 128, 256).}
  \label{fig:model_latency_breakdown}
  \vspace{-0.5cm}
\end{figure}

We observe that \emph{the SLS-family of operators is the largest contributor to latency} in recommendation models especially as batch size, data-level parallelism, increases.
Figure~\ref{fig:model_latency_breakdown} depicts the execution time breakdown per operator with the majority of the time spent executing FC and SLS Caffe2 operators~\cite{arxiv-gupta-19}. 
With a batch size of 8, SLS accounts for 37.2\% and 50.6\% of the total model execution time of RM1-small and RM1-large, respectively. 
Whereas for larger models represented by RM2-small and RM2-large, a more significant portion of the execution time goes into SLS (73.5\%, 68.9\%).
Furthermore, the fraction of time spent on the embedding table operations increases with higher batch-size --- 37.2\% to 61.1\% and 50.6\% to 71.3\% for RM1-small and RM1-large respectively.
Note, the execution time of RM2-large is 3.6$\times$ higher than RM1-large because RM2 comprises a higher number of parallel embedding tables. 
Generally, embedding table sizes are expected to increase further for models used in industry~\cite{tensorDIMM}. 



\vspace{-0.1cm}
\subsection{Roofline Analysis}
\vspace{-0.1cm}

Applying the roofline model~\cite{roofline}, we find \emph{recommendation models lie in the memory bandwidth-constrained region, close to the theoretical roofline performance bound}.
We construct a roofline describing the theoretical limits of the test system described in Section~\ref{sec:method}. We use Intel's Memory Latency Checker (MLC)\footnote{Intel MLC~\cite{MLC} measures the bandwidth from the processor by creating threads that traverse a large memory region in random or sequential stride as fast as possible.} to derive the memory bound. We derive the compute bound by sweeping the number of fused multiply-add (FMA) units in the processor and the operating frequency of the CPU (Turbo mode enabled).

\begin{figure}[t!]
\begin{minipage}[t]{0.44\linewidth}
    \includegraphics[width=\linewidth]{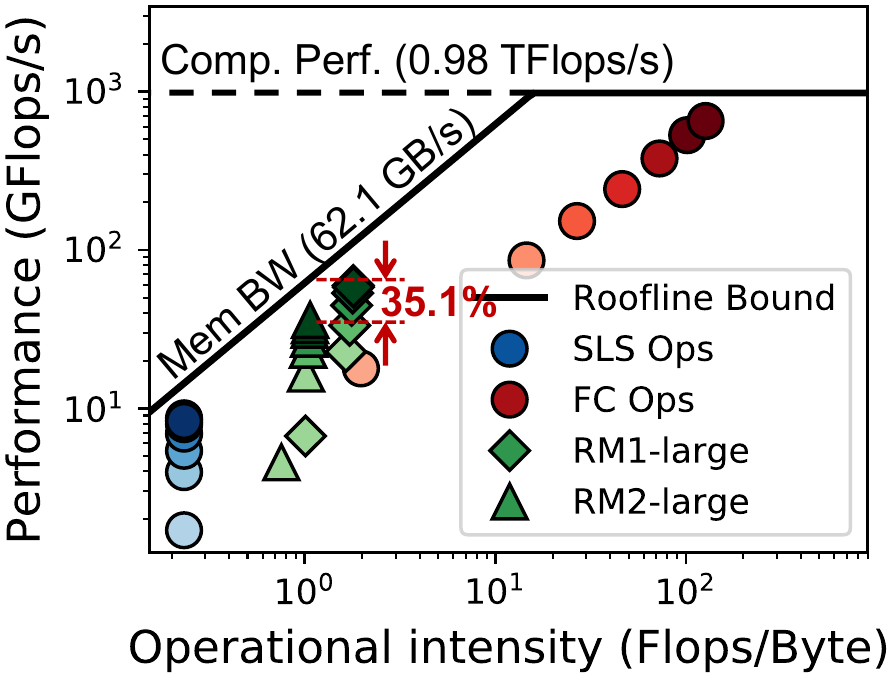}
    \vspace{-0.7cm}
    \caption{Roofline of multi-threaded RM1-large and RM2-large sweeping batch size (1-256). Darker color indicates larger batch.}
    \label{fig:RM2_roofline}
\end{minipage}%
    \hfill%
\begin{minipage}[t]{0.54\linewidth}
    \includegraphics[width=\linewidth]{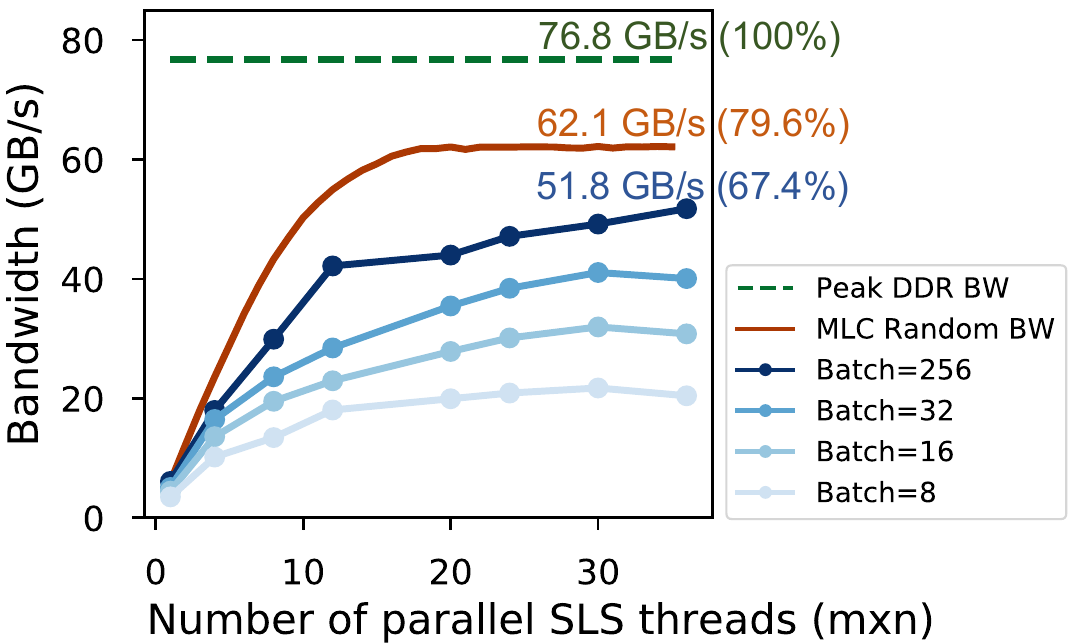}
    \vspace{-0.7cm}
    \caption{Memory bandwidth saturation with increasing number of parallel SLS threads and batch sizes.}
    \label{fig:bandwidth}
\end{minipage} 

\vspace{-0.5cm}
\end{figure}

Figure~\ref{fig:RM2_roofline} presents the roofline data points for the models, RM1 and RM2, as well as their corresponding FC and SLS operators separately. We sweep batch size from 1 to 256 with darker colors indicating a larger batch size. We observe that the SLS operator has low compute but higher memory requirements; the FC portion of the model has higher compute needs; and the combined model is in between.   
SLS has low and fixed operational intensity across batch sizes, as it performs vector lookups and element-wise summation.
FC's operational intensity increases with batch size, as all requests in the batch share the same FC weights, increasing FC data reuse. 
With increasing batch size, the FC operator moves from the region under the memory-bound roofline to the compute-bound region.
For the full model, we find RM1 and RM2 in the memory bound region, as the operational intensity is dominated by the high percentage of SLS operations.
It also reveals that, with increasing batch size, the performance of SLS, as well as the RM1 and RM2 models, is approaching the theoretical performance bound of the system.

More importantly, \textit{our roofline analysis suggests that the performance of the recommendation model is within 35.1\% of the theoretical performance bound and there is little room for further improvement without increasing system memory bandwidth.}
By performing the embedding lookups and pooling operations before crossing the pin-limited memory interface, near-memory processing can exploit higher internal bandwidth of the memory system, 
thus effectively lifting up the roofline and fundamentally improving the memory bandwidth-constrained performance bound.

\vspace{-0.1cm}
\subsection{Memory Bandwidth of Production Configurations} 
\vspace{-0.1cm}

\emph{Executing embedding operations on real systems can saturate memory bandwidth at high model-, operator- and data-level parallelism.} 
Figure~\ref{fig:bandwidth} depicts the memory bandwidth consumption as we increase the number of parallel SLS threads for different batch sizes (blue curves). The green horizontal line represents the ideal peak bandwidth (76.8 GB/s, 4-channel, DDR4-2400) and the red curve is an empirical upper bound measured with Intel MLC~\cite{MLC}.
We observe that memory bandwidth can be easily saturated by embedding operations especially as batch size and the number of threads increase.
In this case, the memory bandwidth saturation point occurs (batch size = 256, thread size = 30) where more than 67.4\% of the available bandwidth is taken up by SLS.
In practice, a higher level of bandwidth saturation beyond this point becomes undesirable as memory latency starts to increase significantly~\cite{intel_optane}.
\textit{What is needed is a system that can perform the Gather-Reduce operation near memory such that only the final output from the pooling returns to the CPU.} 

\vspace{-0.1cm}
\subsection{Embedding Table Locality Analysis}
\label{sec:locality}
\vspace{-0.1cm}

\begin{figure}[t]
  \centering
  \includegraphics[width=\columnwidth]{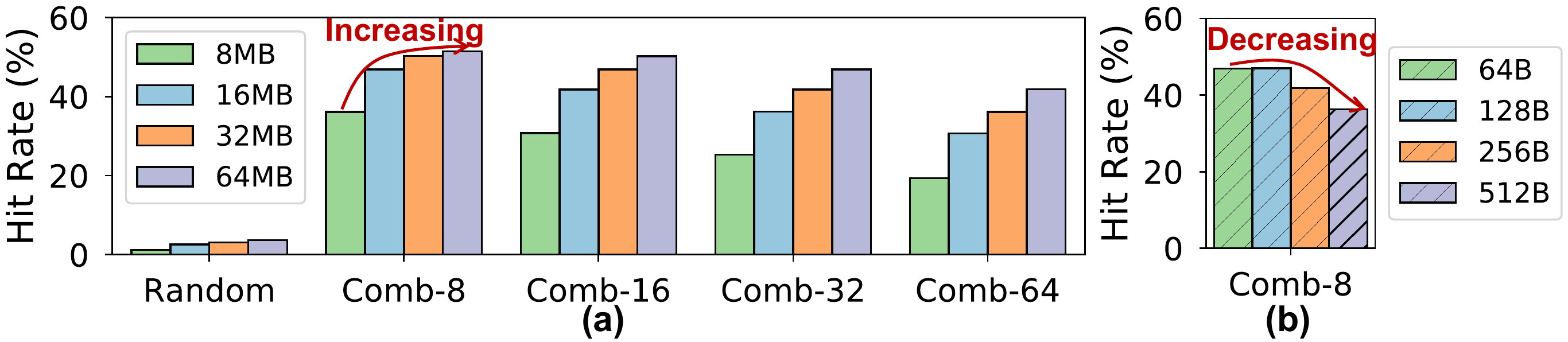}
  \vspace{-0.7cm}
  \caption{(a) Temporal data locality sweeping cache capacity 8-64MB with fixed cacheline size of 64B;
  (b) Spatial data locality sweeping cacheline size 64-512B with fixed cache capacity 16MB.
  }
  \label{fig:sls_lru_locality}
  \vspace{-0.5cm}
\end{figure}

Prior work~\cite{arxiv-gupta-19} has assumed that embedding table lookups are always random, however we show, \emph{for traces from production traffic, there exists modest level of locality mostly due to temporal reuse.}
While recommendation models are limited by memory performance generally, we wanted to study the memory locality to see if caching can improve performance.
We evaluate both a random trace and embedding table (T1-T8) lookup traces from production workloads used by Eisenman et al.~\cite{eisenman2018bandana}. 
In production systems, one recommendation model contains tens of embedding tables and multiple models are co-located on a single machine. 
To mimic the cache behavior of a production system, we simulate the cache hit rate for multiple embedding tables co-located on one machine.
In Figure~\ref{fig:sls_lru_locality}(a), Comb-8 means that 8 embedding tables are running on the machine and the T1-T8 traces (each for a single embedding table) are interleaved for the 8 embedding tables.
For Comb-16, Comb-32 and Comb-64 we multiply the 8 embedding tables 2, 4, and 8 times  on the same machine, which also approximates larger models with 16, 32 and 64 embedding tables.
We use the LRU cache replacement policy and 4-way set associative cache. We  assume each embedding table is stored in a contiguous logical address space and randomly mapped to free physical pages.

To estimate the amount of temporal locality present, we sweep the cache capacity between 8-64MB with fixed cacheline size of 64B.  
In Figure~\ref{fig:sls_lru_locality}(a), the random trace has a low hit rate of $<$5\% representing the worst case locality. We see that the combined simulation of production traces is much higher than random with a hit rate between 20\% and 60\%.
More importantly, hit rate increases as cache size increases. In Section~\ref{sec:co-opt}, we will show how optimizations to \DesName can take advantage of this locality through table-aware packet scheduling and software locality hints from batch profiling. 

Spatial locality can be estimated by sweeping the cacheline size of 64-512B with a fixed cache capacity of 16MB. Figure~\ref{fig:sls_lru_locality}(b) illustrates this sweep for the Comb-8.  
We observe that as the cacheline size increases, in fact, hit rate decreases. In order to isolate the effect of increased conflict misses we run the same experiment on a fully-associative cache and observe similar trends of decreasing hit rate.  
Thus, we conclude that embedding table lookup operations have little spatial locality.

\vspace{-0.1cm}
\section{\DesNameni System Design}
\label{sec:design}
\vspace{-0.1cm}
Considering the unique memory-bounded characteristics and the sparse and irregular access pattern of personalized recommendation, we propose \textit{RecNMP}---a practical and lightweight near-memory processing solution to accelerate the dominated embedding operations.
It is designed to maximize DRAM rank-level parallelism by computing directly and locally on data fetched from concurrently activated ranks.

First, we employ a minimalist style hardware architecture and embed specialized logic units and a rank-level cache to only support the SLS-family inference operators instead of general-purpose computation.
The modified hardware is limited to the buffer chip within a DIMM without requiring any changes to commodity DRAM devices.
Next, the sparse, irregular nature of embedding lookups exerts a high demand on command/address (C/A) bandwidth.
This is addressed by sending a compressed instruction format over the standard memory interface, conforming to the standard DRAM physical pin-outs and timing constraints. Other proposed NMP solutions have employed special NMP instructions without address the C/A limitation of irregular and low spatial locality memory accesses pattern~\cite{chameleon, tensorDIMM}.
We also present a hardware/software (HW/SW) interface for host-NMP coordination by adopting a heterogeneous computing programming model, similar to OpenCL~\cite{opencl}.
Finally, we explore several HW/SW co-optimization techniques--\textit{memory-side caching}, \textit{table-aware scheduling} and \textit{hot entry profiling}--that provide additional performance gains. These approaches leverage our observations from the workload characterization in the previous section.


\vspace{-0.2cm}
\subsection{Hardware Architecture}
\vspace{-0.1cm}

\begin{figure*}[t!]
  \centering
  \includegraphics[width=\textwidth]{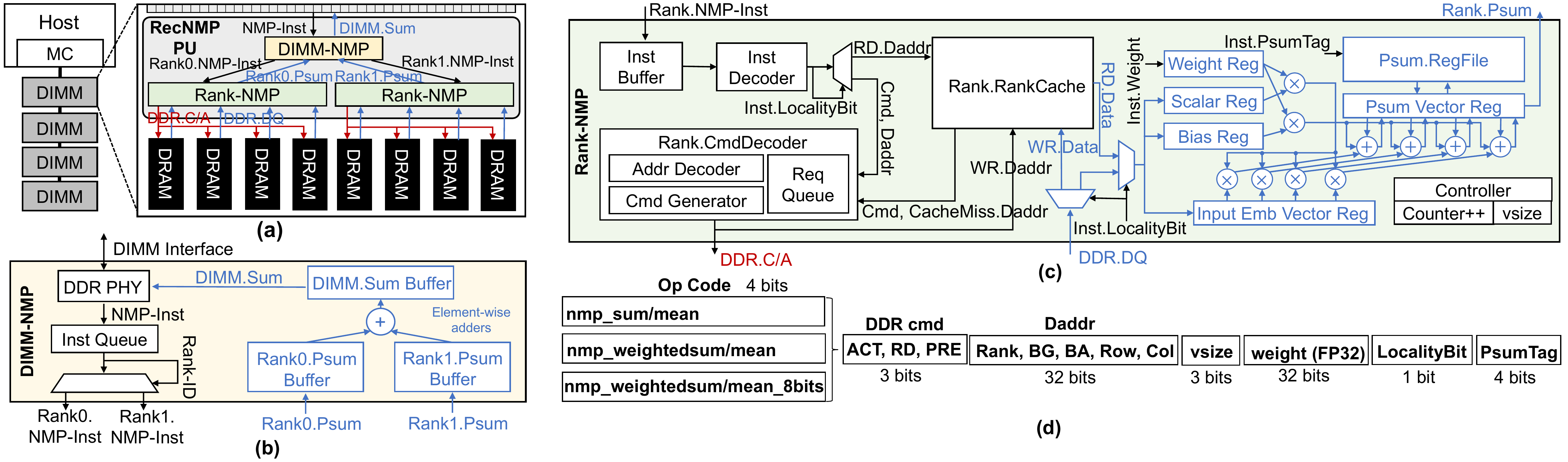}
  \vspace{-0.8cm}
  \caption{(a) Architecture overview of \DesName architecture; (b) DIMM-NMP; (c) Rank-NMP; (d) NMP instruction format.}
  \label{fig:system_overview}
  \vspace{-0.6cm}
\end{figure*}

\textbf{System overview.}
\DesName resides in the buffer chip on the DIMM.
The buffer chip bridges the memory channel interface from the host and the standard DRAM device interface, using data and C/A pins, as illustrated in Figure~\ref{fig:system_overview}(a).
Each buffer chip contains a \DesNameni processing unit (PU) made up of a DIMM-NMP module and multiple rank-NMP modules.
This approach is non-intrusive and scalable, as larger memory capacity can be provided by populating a single memory channel with multiple RecNMP-equipped DIMMs.
Multiple DRR4 channels can also be utilized with software coordination.

The host-side memory controller communicates with a \DesNameni PU by sending customized compressed-format NMP instructions (NMP-Inst) through the conventional memory channel interface; the PU returns the accumulated embedding pooling results (DIMM.Sum) to the host.
Regular DDR4-compatible C/A and data signals (DDR.C/A and DDR.DQ) are decoded by the \DesNameni PU from the NMP-Insts and then sent to all DRAM devices across all parallel ranks in a DIMM.
By placing the logic at rank-level, \DesName is able to issue concurrent requests to the parallel ranks and utilize, for SLS-family operators, the higher internal bandwidth present under one memory channel.
Its effective bandwidth thus aggregates across all the parallel activated ranks. For example, in Figure~\ref{fig:system_overview}(a), a memory configuration of 4 DIMMs$\times$2 ranks per DIMM could achieve $8\times$ higher internal bandwidth.


The DIMM-NMP module first receives a NMP-Inst through DIMM interface and then forwards it to the corresponding rank-NMP module based on the rank address.
The rank-NMPs decode and execute the NMP-Inst to perform the local computation of the embedding vectors concurrently.
We do not confine a SLS operation to a single rank but support aggregation across ranks within the PU. This simplifies the memory layout and increases bandwidth.
DIMM-NMP performs the remaining element-wise accumulation of the partial sum vectors (PSum) from parallel ranks to arrive at the final result (DIMM.Sum). In the same fashion, Psums could be accumulated across multiple \DesNameni PUs with software coordination.
We will next dive into the
design details on the DIMM-NMP and rank-NMP modules. While they are on the same buffer chip, having separate logical modules makes it easy to scale to DIMMs with a different number of ranks.


\textbf{DIMM-NMP Module.}
To dispatch the NMP-Inst received from the DIMM interface, the DIMM-NMP module employs DDR PHY and protocol engine similar to the design of a conventional DIMM buffer chip relaying the DRAM C/A and DQ signals from and to the host-side memory controller.
The instruction is multiplexed to the corresponding ranks based on the Rank-ID as shown in Figure~\ref{fig:system_overview}(b).
DIMM-NMP buffers the Psum vectors accumulated by each rank-NMP in its local registers and performs final summation using an adder tree before sending the final result back to the host via the standard DIMM interface.
Depending on the memory system configuration, the number of ranks within a DIMM can vary, changing the number of inputs to the adder tree.

\textbf{Rank-NMP Module.}
\DesName uses the internal bandwidth on a DIMM to increase the effective bandwidth of embedding table operations, thus the majority of the logic is replicated for each rank. Three crucial functions are performed by the rank-NMP module---translating the NMP-Inst into low-level DDR C/A commands, managing {\it memory-side caching} and local computation of SLS-family operators.
As illustrated in Figure~\ref{fig:system_overview}(c), the NMP-Inst is decoded to control signals and register inputs.
To address C/A bus limitations, all of the DDR commands for a single SLS vector is embedded in one NMP-Inst. Three fields in NMP-Inst (Figure~\ref{fig:system_overview}(d))---DDR cmd (the presence/absence of \{ACT, RD, PRE\} with bit 1/0), vector size (vsize), and DRAM address (Daddr)---determine the DDR command sequence and the burst length. These are fed to the local command decoder (Rank.CmdDecoder) to generate standard DDR-style ACT/RD/PRE commands to communicate with DRAM devices.
The tags are set at runtime by the host-side memory controller based on the relative physical address location of consecutive embedding accesses. This keeps the CmdDecoder in rank-NMP lightweight, as the host-side memory controller has performed the heavy-lifting tasks of request reordering, arbitration, and clock and refresh signal generation.
If a 128B vector (vsize=2) requires ACT/PRE from a row buffer miss, the command sequence to DRAM devices for the NMP-Inst is \{PRE, ACT Row, RD Col, RD Col+8\} decoded from \{ACT, RD, PRE\} and vsize tags.

Our locality analysis in Section~\ref{sec:char} shows that the modest temporal locality within some embedding tables as vectors are reused.
The operands of each SLS-family operator vary so caching the final result in the DIMM or CPU will be ineffective.
We incorporate a memory-side cache (RankCache) in each rank-NMP module to exploit the embedding vectors reuse.
The RankCache in \DesName takes hints from the LocalityBit in the NMP-Inst to determine whether an embedding vector should be cached or bypassed.
The detailed method to generate the LocalityBit hint through hot entry profiling will be explained in Section~\ref{sec:co-opt}.
Entries in RankCache are tagged by the DRAM address field (Daddr).
If the LocalityBit in the NMP-Inst indicates low locality,
the memory request bypasses the RankCache and is forwarded to Rank.CmdDecoder to initiate a DRAM read. Embedding tables are read-only during inference, so this optimization does not impact correctness.

The datapath in the rank-NMP module supports a range of SLS-family operators.
The embedding vectors returned by the RankCache or DRAM devices are loaded to the input embedding vector registers.
For weighted sum computation, the weight registers are populated by the weight fields from the NMP-Inst.
For quantized operators such as the SLS-8bits operator, the dequantized parameters $Scalar$ and $Bias$ are stored with the embedding vectors and can be fetched from memory to load to the Scalar and Bias registers.
The Weight and Scalar/Bias registers are set to be 1 and 1/0 during execution of non-weighted and non-quantized SLS operators.
The PsumTag decoded from the NMP-Inst is used to identify the embedding vectors belonging to the same pooling operations, as multiple poolings in one batch for one embedding table could be served in parallel.
The controller counter, vector size register,
and final sum registers in the both the DIMM-NMP and rank-NMP modules are all memory-mapped, easily accessible and configurable by the host CPU.

\vspace{-0.1cm}
\subsection{C/A Bandwidth Expansion}
\label{sec:mem_config}
\vspace{-0.1cm}

Although the theoretical aggregated internal bandwidth of \DesName scales linearly with the number of ranks per channel, in practice, the number of concurrently activated ranks is limited by the C/A bandwidth.
Due to frequent row buffer misses/conflicts from low spatial locality, accessing the embedding table entries in memory requires a large number of ACT and PRE commands.
The reason is that the probability of accessing two embedding vectors in the same row is quite low, as spatial locality only exists in continuous DRAM data burst of one embedding vector.
In production, embedding vector size ranges from 64B to 256B with low spatial locality, resulting in consecutive row buffer hits in the narrow range of 0 to 3.

\begin{figure}[t!]
  \centering
  \includegraphics[width=\columnwidth]{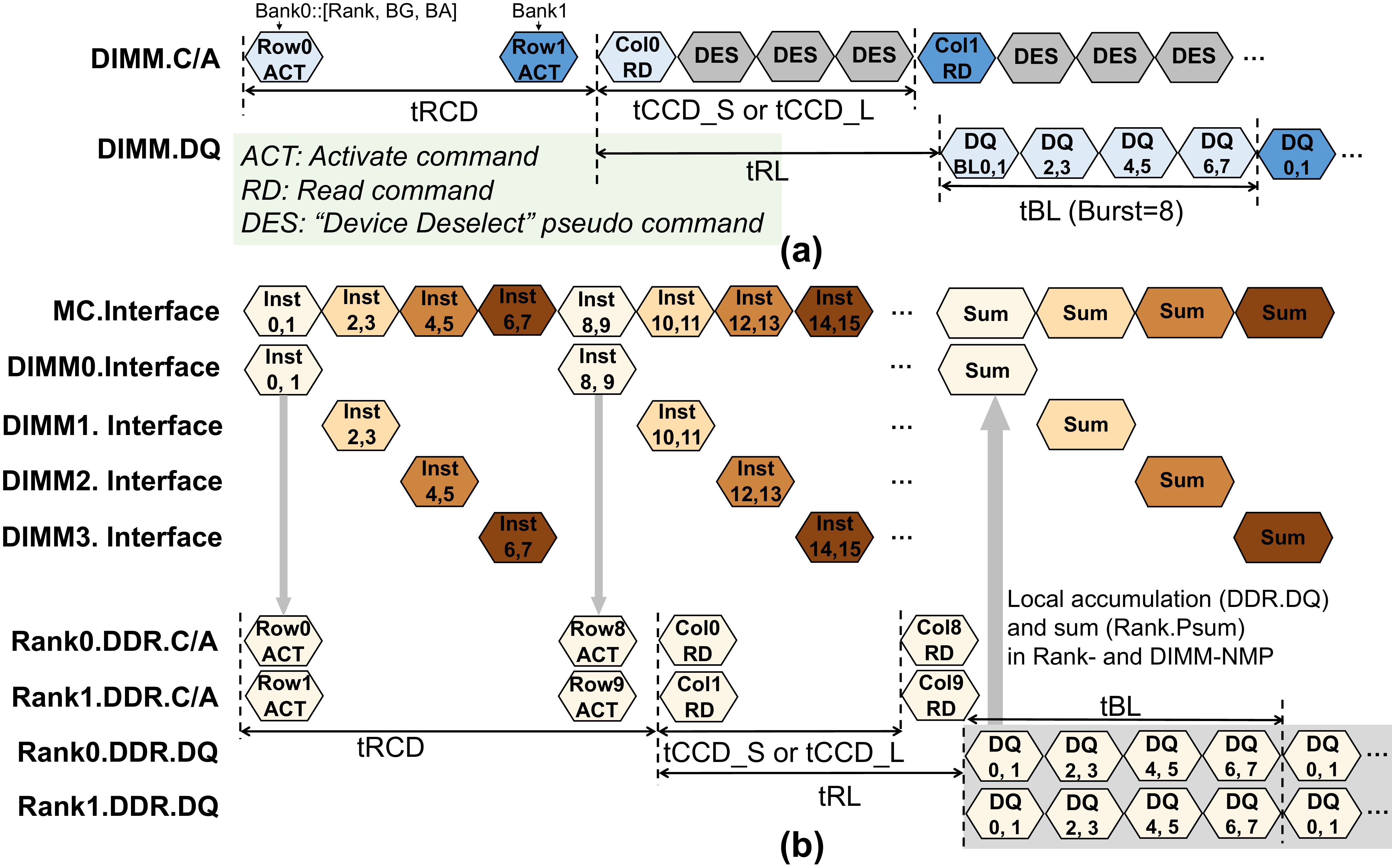}
  \vspace{-0.7cm}
  \caption{Timing diagram of (a) ideal DRAM bank interleaving read operations; (b) The proposed \DesName concurrent rank activation.
  }
  \label{fig:timing}
  \vspace{-0.6cm}
\end{figure}

To fully understand the C/A bandwidth limitation, we analyze the worst-case scenario when the embedding vector size is 64B.
A typical timing diagram is presented in Figure~\ref{fig:timing}(a). It shows an ideal sequence of bank-interleaved DRAM reads that could achieve one consecutive data burst.
In this burst mode, the \emph{ACT} command first sets the row address.
Then the \emph{RD} command is sent accompanied by the column address.
After $t_{RL}$ DRAM cycles, the first set of two 64-bit data (DQ0 and DQ1) appear on the data bus.
The burst mode lasts for 4 DRAM cycles (burst length = 8) and transmits a total of 64B on the DQ pins at both rising and falling edges of the clock signal.
Modern memory systems employ bank interleaving, therefore in the next burst cycle (4 DRAM cycles), data from a different bank can be accessed in a sequential manner.
In this ideal bank interleaving case, every 64B data transfer takes 4 DRAM cycles and requires 3 DDR commands (ACT/RD/PRE) to be sent over the DIMM C/A interface, this consumes 75\% of the C/A bandwidth.
Activating more than one bank concurrently would require issuing more DDR commands, thus completely exhausting the available C/A bandwidth of conventional memory interface.

To overcome C/A bandwidth limitation, we propose a customized NMP-Inst with a compressed format of DDR commands to be transmitted from memory controller to \DesNameni PUs.
Figure~\ref{fig:timing}(b) illustrates the timing diagram of interleaving NMP-Inst to a 4 DIMMs $\times$ 2 Ranks per DIMM memory configuration.
Eight NMP-Insts can be transferred between memory controller and DIMMs interfaces in 4 DRAM data burst cycles on double data rate.
In low spatial locality case (64B embedding vector and one NMP-Inst per vector) and ideal bank interleaving, we could potentially activate 8 parallel ranks to perform 8${\times}$64B lookups concurrently in 4 DRAM data burst cycles.
Although customized instructions have been proposed before~\cite{nda-kim, chameleon, tensorDIMM}, our solution is the first one to directly deal with the C/A bandwidth limitation using DDR command compression that enables up to $8\times$ bandwidth expansion for small-sized embedding vectors (i.e. 64B) with low spatial locality.
Higher expansion ratio can be achieved with larger vector size.

\vspace{-0.25cm}
\subsection{Programming Model and Execution Flow}
\vspace{-0.1cm}

Like previous NMP designs~\cite{chameleon,pim-NN-training}, \DesName adopts a heterogeneous computing programming model (e.g. OpenCL),
where the application is divided into host calls running on the CPU and NMP kernels being offloaded to RecNMP PUs.
NMP kernels are compiled into packets of NMP-Insts and transmitted to each memory channel over the DIMM interface to \DesName PUs.
Results of NMP kernels are then transmitted back to the host CPU.
In Figure~\ref{fig:system_overview}(d), each 79-bit NMP-Inst contains distinctive fields that are associated with different parameters in an embedding operation, locality hint bit (LocalityBit) and pooling tags (PsumTag) passed between the HW/SW interface.
The proposed NMP-Inst format can fit within the standard 84-pin C/A and DQ interface.

\begin{figure}[t!]
  \centering
  \includegraphics[width=\columnwidth]{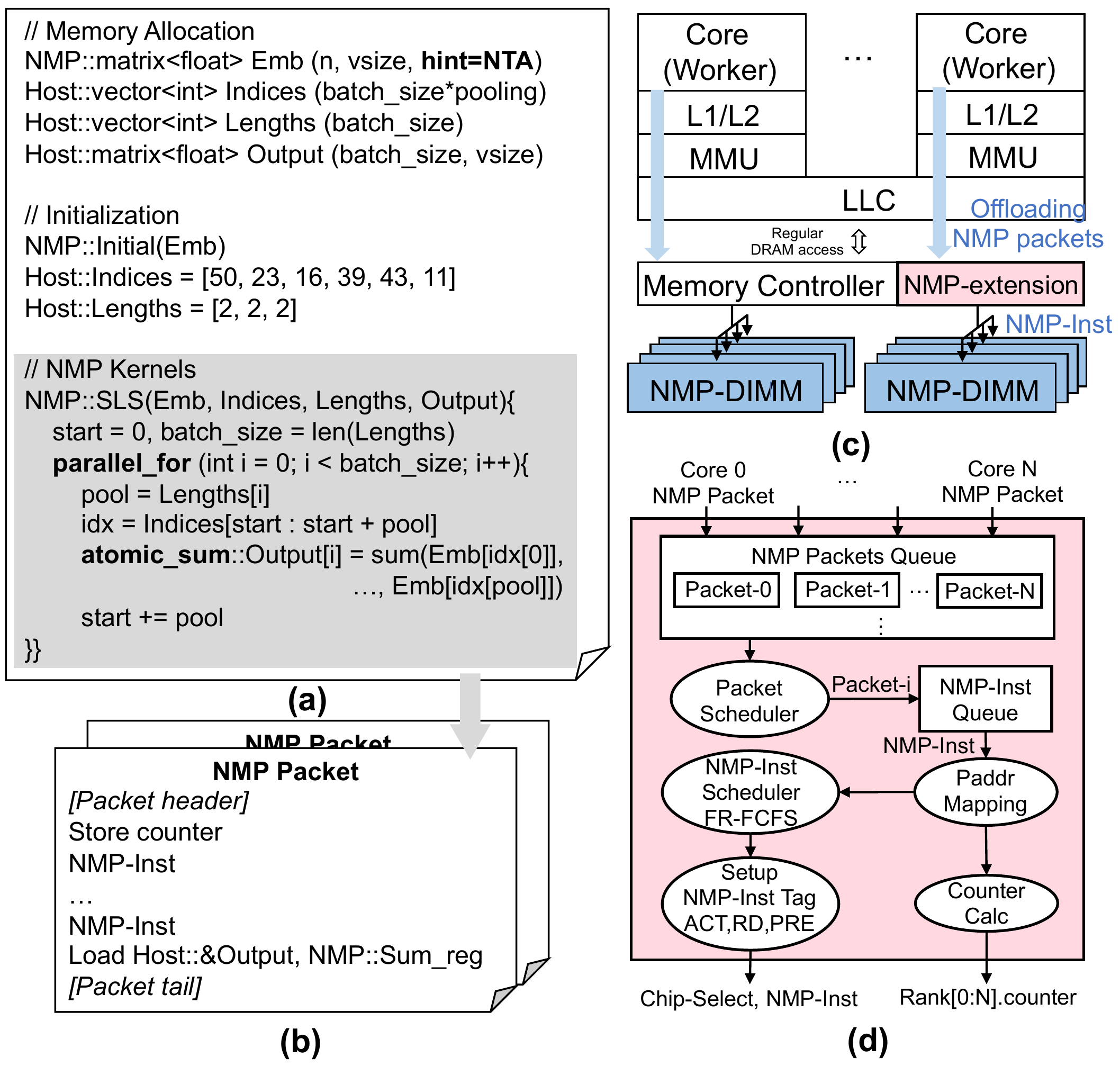}
  \vspace{-0.7cm}
  \caption{(a) \DesName SLS example code; (b) NMP packet; (c) NMP kernel offloading; (d) NMP-enabled memory controller.}
  \label{fig:offload_flow}
  \vspace{-0.7cm}
\end{figure}

Using a simple SLS function call in Figure~\ref{fig:offload_flow}(a) as an example, we walk through the execution flow of the proposed \DesName programming model.
First, memory is allocated for SLS input and output data, and is marked up as either Host (cacheable) or NMP (non-cacheable) regions to simplify memory coherence between the host and \textit{RecNMP}.
Variables containing host visible data, such as the two arrays \emph{Indices and Lengths}, are initialized and loaded by the host and are cachable in the host CPU's cache hierarchy.
The embedding table (Emb) in memory is initialized by the host as a host non-cacheable NMP region using a non-temporal hint (NTA)~\cite{nta}.

Next, the code segment
marked as a NMP kernel is compiled to packets of NMP-Insts (Figure~\ref{fig:offload_flow}(b)).
A single SLS NMP kernel containing one batch of embedding poolings can be split into multiple NMP packets, with each packet having one or more pooling operations.
The NMP-Insts belonging to different embedding poolings in one NMP packet are tagged by PsumTag, and the maximum number of poolings in one packet is determined by the number of bits of the PsumTag.
We use a 4-bit PsumTag in our design.
At runtime, the NMP kernel is launched by the host with special hardware/driver support to handle NMP packet offloading; access to the memory management unit (MMU) to request memory for NMP operations; and the virtual memory system for logical-to-physical addresses translation (Figure~\ref{fig:offload_flow}(c)).
The offloaded NMP packets bypass L1/L2 and eventually arrive at the host-side memory controller with an NMP extension.
To avoid scheduling the NMP packets out-of-order based on FR-FCFS policy, the NMP extension of the memory controller includes extra scheduling and arbitration logic.

As illustrated in Figure~\ref{fig:offload_flow}(d), the memory controller with the NMP extension receives concurrent NMP packets from parallel execution of multiple host cores, which are stored in a queue.
Once scheduled, each NMP packet is decoded into queued NMP-Insts.
Physical-to-DRAM address mapping is then performed and a FR-FCFS scheduler reorders the NMP-Insts within a packet only and not between packets.
Instead of sending direct DDR commands, ACT/RD/PRE actions are compressed into the 3-bit DDR\_cmd field in the NMP-Inst.
The host-side memory controller also calculates the correct accumulation counter value to configure the memory-mapped control registers in the \DesName PU.
Finally, after the completion of all the counter-controlled local computation inside the \DesName PU for one NMP packet, the final summed result is transmitted over the DIMM interface and returned to the \emph{Output} cacheable memory region visible to the CPU.

\vspace{-0.2cm}
\subsection{HW/SW Co-optimization}
\label{sec:co-opt}
\vspace{-0.1cm}

Our locality analysis of production recommendation traffic in Section~\ref{sec:locality} illustrates intrinsic temporal reuse opportunities in embedding table lookups.
We propose memory-side caches (RankCache) inside rank-NMP modules.
To extract more performance from memory-side caching, we explore two additional HW/SW co-optimization techniques.
This locality-aware optimization results in 33.7\% memory latency improvement and 45.8\% memory access energy saving (detailed performance benefits will be presented in Section V).

\begin{figure}[t]
  \centering
  \includegraphics[width=\columnwidth]{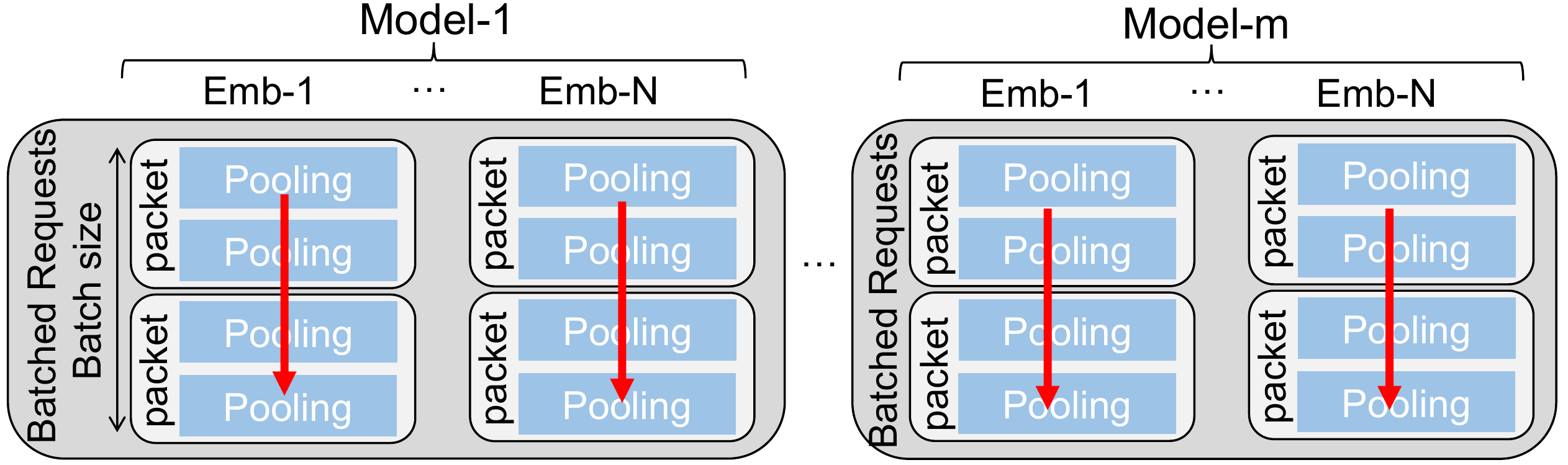}
  \vspace{-0.7cm}
  \caption{
  NMP packet scheduling scheme that prioritizes batch of single table.}
  \label{fig:intra_mini_batch_opt}
  \vspace{-0.5cm}
\end{figure}

\begin{figure}[t]
  \centering
  \includegraphics[width=\columnwidth]{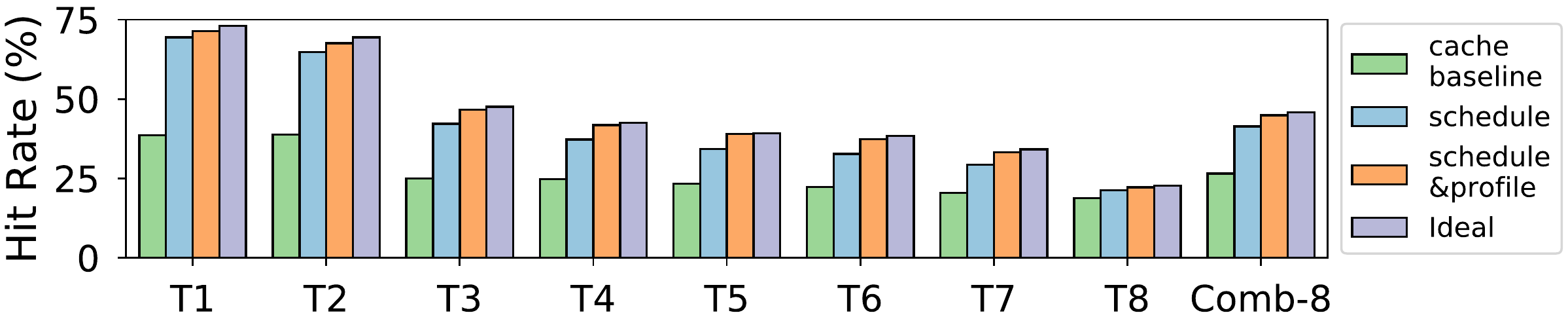}
  \vspace{-0.7cm}
  \caption{Hit rate of 1MB cache without optimization, with table-aware packet scheduling optimization, with both table-aware packet scheduling and hot entry profiling optimization, and ideal case without interference.
  }
  \label{fig:opt_hit_rate}
  \vspace{-0.5cm}
\end{figure}

First, to preserve the intrinsic locality from embedding lookups residing in one table, we propose to prioritize scheduling NMP packets from a single batch requests to the same embedding table together --  {\it table-aware packet scheduling}.
In production workloads, the memory controller receives NMP packets from parallel SLS threads with equal scheduling priority.
The intra-embedding table temporal locality is not easily retained because of the interference from lookup operations of multiple embedding tables.
This locality can be further degraded when multiple recommendation models are co-located.
Therefore, as illustrated in Figure~\ref{fig:intra_mini_batch_opt}, we propose an optimized table-aware NMP packet scheduling strategy to exploit the intrinsic temporal locality within a batch of requests by ordering packets from the same embedding table in one batch first, allowing the embedding vectors to be fetched together, thereby retaining the temporal locality.
SLS operators access separate embedding tables as running in parallel threads, the mechanics of our implementation comes from the thread-level memory scheduler~\cite{mem-schedule}.

Next, we propose another optimization technique -- {\it hot entry profiling}, built on top of the observation that a small subset of embedding entries exhibit relatively higher reuse characteristics.
We profile the vector of indices used for embedding table lookup in an NMP kernel and mark the entries with high locality by explicitly annotating NMP-Insts with a \emph{LocalityBit}.
NMP-Inst with LocalityBit set will be cached in the RankCache; otherwise, the request will bypass the RankCache.
This hot entry profiling step can be performed before model inference and issuing SLS requests and only costs $<$2\% of total end-to-end execution time.
We profile the indices of each incoming batch of embedding lookups and set LocalityBit if the vectors are accessed $>t$ times within the batch.
Infrequent ($<t$ times) vectors will bypass the RankCache and are read directly from the DRAM devices.
We sweep the threshold $t$ and pick the value with the highest cache hit rate to use in our simulation.
This hot entry profiling optimization reduces cache contention and evictions caused by the less-frequent entries in the RankCache.

Figure~\ref{fig:opt_hit_rate} depicts the hit rate improvement when the different optimizations are applied.
Comb-8 indicates the overall hit rate at model level of 8 embedding tables (T1-T8).
To gain more insights, we investigate the hit rate of embedding tables (T1 to T8) in Comb-8.
The ideal bar indicates the theoretical hit rate with an infinitely sized cache.
With the proposed co-optimization, the measured hit rate closely approaches the ideal case across the individual embedding tables, even for the trace with limited locality (T8), illustrating the proposed technique can effectively retain embedding vectors with high likelihood of reuse in RankCache.
\vspace{-0.2cm}

\section{Experimental Methodology}
\label{sec:method}
\begin{figure}[t!]
  \centering
  \includegraphics[width=\columnwidth]{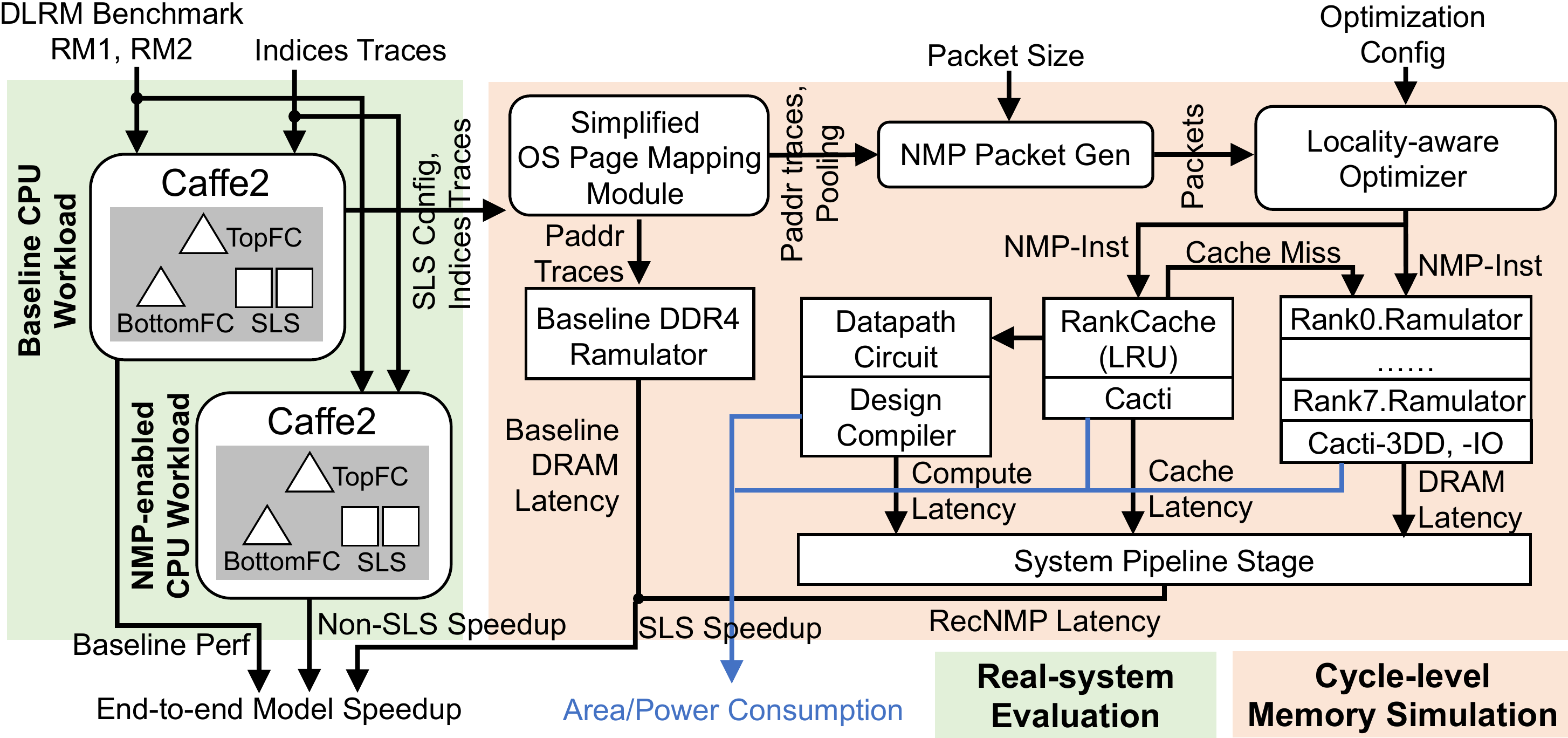}
  \vspace{-0.7cm}
  \caption{\DesName experimental methodology.}
  \label{fig:simulation_setup}
  \vspace{-0.5cm}
\end{figure}

\vspace{-0.1cm}
Our experimental setup combines real-system evaluations with cycle-level memory simulations, as presented in Figure \ref{fig:simulation_setup}. 
For real-system evaluations, we run production-scale recommendation models on server-class CPUs found in the data center. 
This allows us to measure the impact of accelerating embedding operations as well as the side-effect of improved memory performance of FC operations on end-to-end models.
Cycle-level memory simulations allow us to evaluate the design tradeoffs when DRAM systems are augmented with \textit{RecNMP}. Table~\ref{tab:sys_config} summarizes the parameters and configurations used in the experiments.
We ran experiments on an 18-core Intel Skylake with DDR4 memory. The DRAM simulation used standard DDR4 timing from a Micron datasheet~\cite{ddr4-datasheet}.

\textbf{Real-system evaluation.}
We configured the DRLM benchmark with the same model parameters and traces in Figure~\ref{fig:sparsenn}(b) and Section~\ref{sec:char}. 
The workload characterization (Section~\ref{sec:char}) and real-system experiments (Section~\ref{sec:eval}) are performed on single socket Intel Skylake servers, specifications in Table~\ref{tab:sys_config}.

\begin{table}[t!]
\caption{System Parameters and Configurations}
\vspace{-0.2cm}
\label{tab:sys_config}
\centering
\begin{tabular}{l|c|l|c}

\hline\hline
\multicolumn{4}{c}{\textbf{Real-system Configurations}}\\

\hline
Processor & 18 cores, 1.6 GHz & L1I/D & 32 KB\\
\hline
L2 cache & 1 MB & LLC & 24.75 MB \\
\hline
DRAM & \multicolumn{3}{c}{\begin{tabular}[c]{@{}c@{}}DDR4-2400MHz 8Gb $\times$8, 64 GB, \\ 4 Channels $\times$ 1 DIMM $\times$ 2 Ranks, FR-FCFS\\ 32-entry RD/WR queue, Open policy,\\ Intel Skylake address mapping~\cite{skladdr}\end{tabular}} \\

\hline\hline
\multicolumn{4}{c}{\textbf{DRAM Timing Parameters}} \\
\hline
\multicolumn{4}{c}{\begin{tabular}[c]{@{}c@{}} tRC=55, tRCD=16, tCL=16, tRP=16, tBL=4\\ tCCD\_S=4, tCCD\_L=6, tRRD\_S=4, tRRD\_L=6, tFAW=26\end{tabular}}\\

\hline\hline
\multicolumn{4}{c}{\textbf{Latency/Energy Parameters}} \\
\hline

\multicolumn{4}{c}{\begin{tabular}[c]{@{}c@{}} DDR Activate = 2.1nJ, DDR RD/WR = 14pJ/b, Off-chip IO = 22pJ/b\\
RankCache RD/WR = 1 cycle, 50pJ/access,\\ FP32 adder = 3 cycles, 7.89pJ/Op, FP32 mult = 4 cycles, 25.2pJ/Op\\\end{tabular}}\\
\hline\hline

\end{tabular}
\vspace{-0.7cm}
\end{table}


\textbf{Cycle-level memory simulation.}
We build the \DesName cycle-level simulation framework with four main components: (1) physical addresses mapping module; (2) packet generator; (3) locality-aware optimizer; and (4) a cycle-accurate model of a \DesNameni PU consisting of DRAM devices, RankCache, arithmetic and control logic.
We use Ramulator \cite{ramulator} to conduct cycle-level evaluations of DDR4 devices. On top of Ramulator, we build a cycle-accurate LRU cache simulator for RankCache and model of the 4-stage pipeline in the rank-NMP module.
Cacti~\cite{cacti} is used to estimate the access latency and area/energy of RankCache.
The hardware implementation used to estimate the latency, area and power of the arithmetic logic is built from Synopsys Design Compiler with a commercial 40nm technology library.
To estimate the DIMM energy, we use Cacti-3DD~\cite{cacti-3dd} for DRAM devices and Cacti-IO~\cite{cacti-io} for off-chip I/O at the DIMM level.

During simulation we emulate the scheduling packet generation steps taken by the software stack and the memory controller. First, we apply a standard page mapping method~\cite{page_map} to generate the physical addresses from a trace of embedding lookups by assuming the OS randomly selects free physical pages for each logical page frame.
This physical address trace is fed to Ramulator to estimate baseline memory latency.
For \DesName workloads, the packet generator divides the physical address trace into packets of NMP-Insts that are sent to the cycle-accurate model. 
Next, the when evaluating systems with HW/SW co-optimizations, the locality-aware optimizer performs table-aware packet scheduling and hot entry profiling and decides the sequence of NMP-Insts. \DesName activate all memory ranks in parallel and traditional DRAM bank-interleaving is also used.
For each NMP packet, performance is determined by the slowest rank that receives the heaviest memory request load. 
Rank-NMP and DIMM-NMP logic units are pipelined to hide the latency of memory read operations.
The total latency of \DesName includes extra DRAM cycles during initialization to configure the accumulation counter and the vector size register and a cycle in the final stage to transfer the sum to the host. The latency, in DRAM cycles, of the major components including RankCache, rank-NMP logic performing  weighted-partial sum and final sum are in Table~\ref{tab:sys_config}. 

\section{Evaluation Results} \label{sec:eval}
\vspace{-0.05cm}

This section presents a quantitative evaluation of \DesName and shows it accelerates end-to-end personalized recommendation inference by up to $4.2\times$.
We first present the latency improvement of the offloaded SLS operators on a baseline system before analysing different optimizations including placement with page coloring, memory-side caching, table-aware packet scheduling and hot-entry profiling. We compare \DesName with state-of-the-art NMP systems TensorDIMM and Chameleon~\cite{tensorDIMM,chameleon}.
We also analyze the effect of \DesName on co-located FC operators.
Finally, an end-to-end evaluation of throughput improvement and energy savings at the model level and the area/power overhead is presented.

\vspace{-0.3cm}
\subsection{SLS Operator Speedup}
\vspace{-0cm}

\begin{figure}[t!]
  \centering
  \includegraphics[width=\columnwidth]{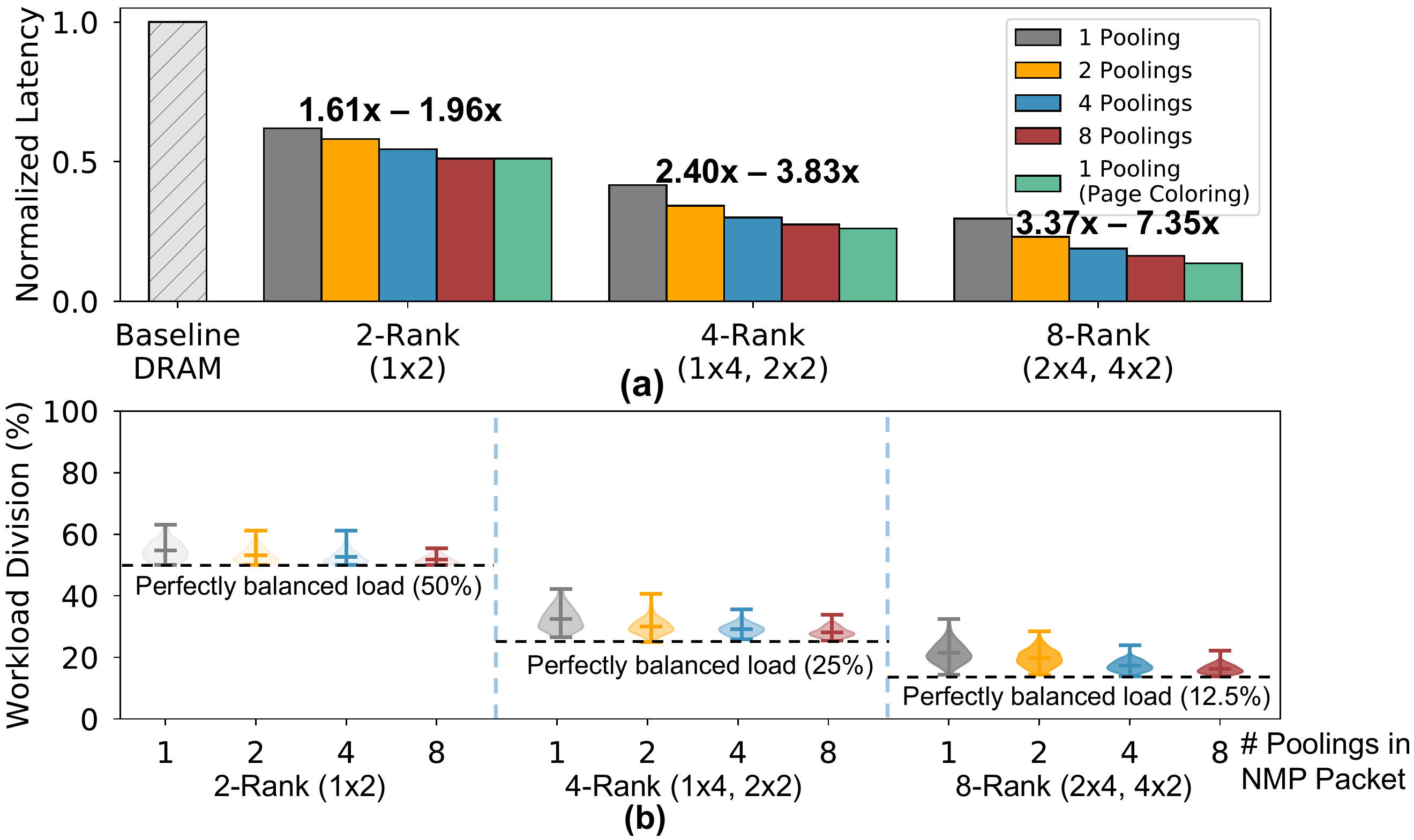}
  \vspace{-0.7cm}
  \caption{(a) Normalized latency of RecNMP-base to the baseline DRAM with different memory configuration (DIMM x Rank) and NMP packet size; (b) Distribution of rank-level load imbalance for 2-, 4-, and 8-rank systems.}
  \label{fig:sls_speedup_no_cache}
  \vspace{-0.6cm}
\end{figure}

In theory, because \DesName exploits rank-level parallelism, speedup will scale linearly with the number of ranks and number of DIMMs in a system.
Therefore, we choose four memory channel configurations (\# of DIMMs $\times$ \# of ranks per DIMM) that correspond to $1\times2$, $1\times4$ and $2\times2$, and $4\times2$ to demonstrate a range of system implementations. 


\textbf{Basic \DesName design without RankCache.} 
We start by evaluating \DesName without a RankCache (RecNMP-base).
In addition to varying the DIMM/rank configuration, we sweep the number of poolings in one NMP packet, where one pooling, in DLRM, is the sum of 80 embedding vectors.
In Figure~\ref{fig:sls_speedup_no_cache}(a), we find 1) SLS latency indeed scales linearly as we increase the number of active ranks in a channel; 2) latency also decreases when there are more pooling operations in an NMP packet.
The variation we observe, as well as the performance gap observed between the actual speedup and the theoretical speedup ($2\times$ for 2-rank, $4\times$ for 4-rank, and $8\times$ for 8-rank systems) is caused by the uneven distribution of embedding lookups across the ranks.
As the ranks operate in parallel, the latency of the SLS operation is determined by the slowest rank, the rank that runs more embedding lookups.
Figure~\ref{fig:sls_speedup_no_cache}(b) shows the statistical distribution of fraction of the work run on the slowest rank. 
When the NMP packet has fewer NMP-Insts, the workload distributes more unevenly, resulting in a longer tail that degrades average speedup.

To address the load imbalance, we experiment with software methods to allocate an entire embedding table to the same rank.
One software approach to perform such data layout optimization is page coloring~\cite{page-coloring}.
As indicated in Figure~\ref{fig:sls_speedup_no_cache}(a), page coloring could achieve 1.96$\times$, 3.83$\times$ and 7.35$\times$ speedup in 2-rank, 4-rank and 8-rank system compared with the DRAM baseline.
The specific page coloring mechanism can be implemented in the operating system by assigning a fixed color to the page frames used by an individual embedding table. The virtual memory system would need to be aware of the DRAM configuration to allocate pages of the same color to physical addresses that map to the same rank.
This data layout optimization can lead to near-ideal speedup, but it requires maintaining high model- and task-level parallelism such that multiple NMP packets from different SLS operators can be issued simultaneously to all the available ranks.

\begin{figure}[t!]
  \centering
  \includegraphics[width=\columnwidth]{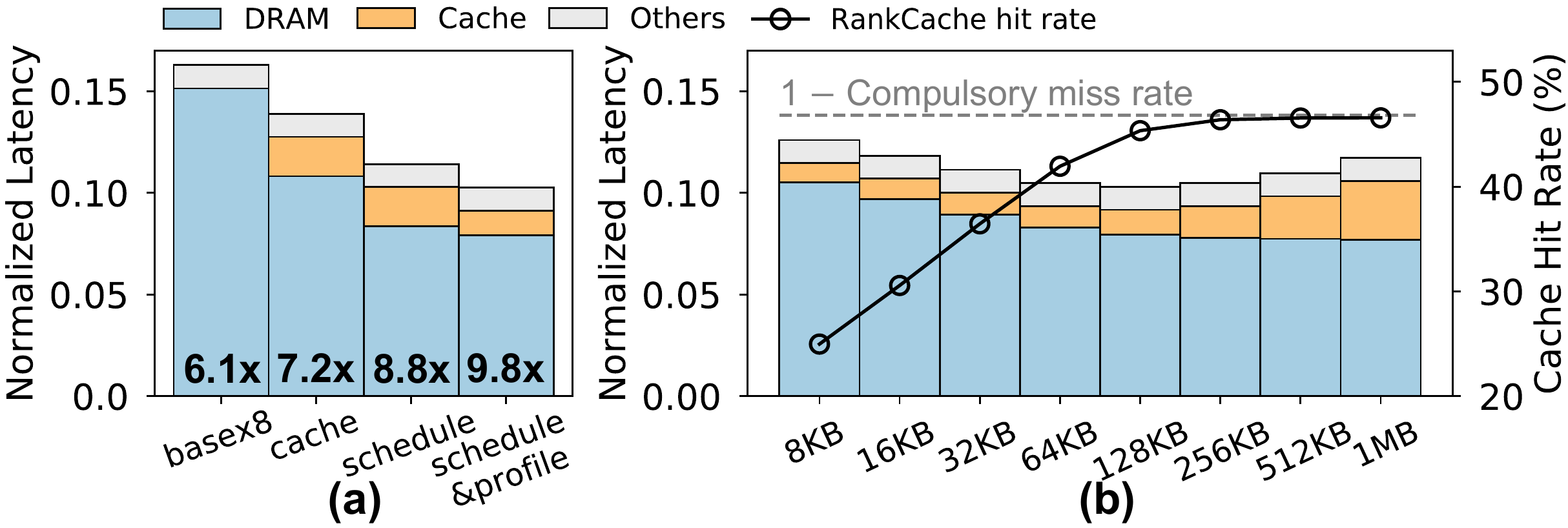}
  \vspace{-0.7cm}
  \caption{(a) Normalized latency of RecNMP-cache and RecNMP-opt with schedule and hot-entry profile optimization to the baseline DRAM system; (b) Cache size sweep effects in RecNMP-opt.}
  \label{fig:sls_speedup_cache_opt}
  \vspace{-0.5cm}
\end{figure}

\textbf{\DesName with RankCache and co-optimization.}
Memory-side caching at the rank-level with table-aware packet scheduling and hot entry profiling is one of the key features of \textit{RecNMP}; these optimizations are described in Section~\ref{sec:co-opt}.
Figure~\ref{fig:sls_speedup_cache_opt}(a) depicts the performance benefits (i.e. latency reduction) enabled by applying different optimization techniques: 1) adding a RankCache, 2) scheduling accesses to the same table together, 3) adding a cachability hint bit from software.
Using a configuration with 8-ranks 8 poolings per packet, we observe 14.2\% latency improvement by adding a 128KB RankCache and an additional 15.4\% improvement by prioritizing the scheduling of NMP packets from the same table and batch. 
In the final combined optimization, \emph{schedule + profile}, we pass cacheability hint after profiling the indices in the batch which reduces cache contention and allows low-locality requests not marked for caching to bypass the RankCache, 
delivering another 7.4\% improvement.
The total memory latency speedup achieved by offloading SLS to an optimized design (RecNMP-opt)  is $9.8\times$.

In Figure~\ref{fig:sls_speedup_cache_opt}(b), we sweep RankCache capacity from 8KB to 1MB and display how cache size affects the normalized latency and cache hit rate.
When RankCache is small (e.g. 8KB), the low cache hit rate (e.g. 24.9\%) leads to high DRAM access latency.
The performance reaches the optimal design point at 128KB.
Further increase of cache size has marginal improvement on hit rate, since it already reaches the compulsory limit in the trace.
Yet it incurs longer cache access latency and degrades overall performance.

\begin{figure}[t!]
  \centering
  \includegraphics[width=\columnwidth]{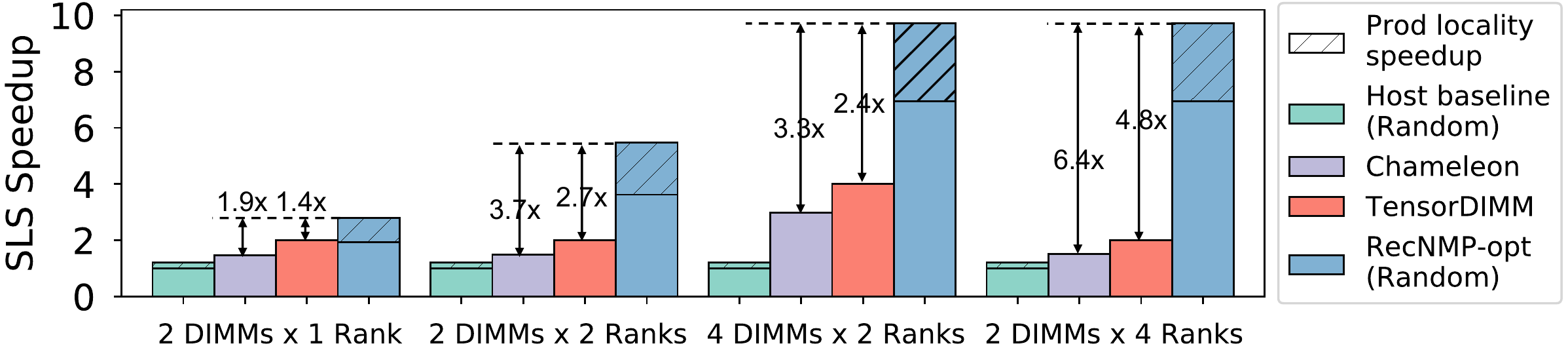}
  \vspace{-0.7cm}
  \caption{Comparison between Host baseline, RecNMP-opt, TensorDIMM~\cite{tensorDIMM} and Chameleon~\cite{chameleon} with both random and production traces}
  \label{fig:sls_speedup_tensordimm}
  \vspace{-0.6cm}
\end{figure}

\textbf{Performance comparison.}
We compare \DesName with state-of-the-art NMP designs such as Chameleon~\cite{chameleon} and TensorDIMM~\cite{tensorDIMM}.
Both are DIMM-based near-memory processing solutions.
TensorDIMM scales the embedding operation performance linearly with the number of parallel DIMMs. 
Since non-SLS operators are accelerated by GPUs in TensorDIMM, which is orthogonal to near-memory acceleration techniques, we only compare its memory latency speedup with \textit{RecNMP}.
Chameleon does not directly support embedding operations. 
We estimate its performance of Chameleon by simulating the temporal and spatial multiplexed C/A and DQ timing of Chameleon's NDA accelerators.
In Figure~\ref{fig:sls_speedup_tensordimm}, as \DesName exploits rank-level parallelism, its performance scales when either the number of DIMMs and ranks increase, whereas Chameleon and TensorDIMM only scale by increasing the number of DIMMs.
This is evident as we sweep the memory channel configuration.
When we increase the number of ranks per-DIMM, \DesName can deliver 3.3-6.4$\times$ and 2.4-4.8$\times$ better performance than Chameleon and TensorDIMM. 

It is also worth noting that \DesName has performance advantages ($1.9\times$ and $1.4\times$) even in configurations with one rank per DIMM, thanks to the memory-side caching, table-aware packet scheduling, and hot-entry profiling optimization techniques.
Neither Chameleon nor TensorDIMM includes a memory-side cache to explicitly take advantage of the available locality in the memory access patterns, hence their performance, with respect to memory latency, is agnostic to traces with different amounts of data reuse.
In contrast, \DesName design can extract 40\% more performance (shown as shaded) from production traces when compared to fully random traces. 


\vspace{-0.2cm}
\subsection{FC Operator Speedup}
\label{sec:FC-exp}
\vspace{-0.1cm}

Although \DesName is designed to accelerate the execution of SLS operators, it can also improve FC performance by alleviating cache contention caused by model co-location. As the degree of data-level parallelism increases, the FC weights brought into the cache hierarchy have higher reuse, normally resulting in fewer cache misses.
However, when co-located with other models, reusable FC data are often evicted early from the cache by SLS data, causing performance degradation.

\begin{figure}[t!]
  \centering
  \includegraphics[width=\columnwidth]{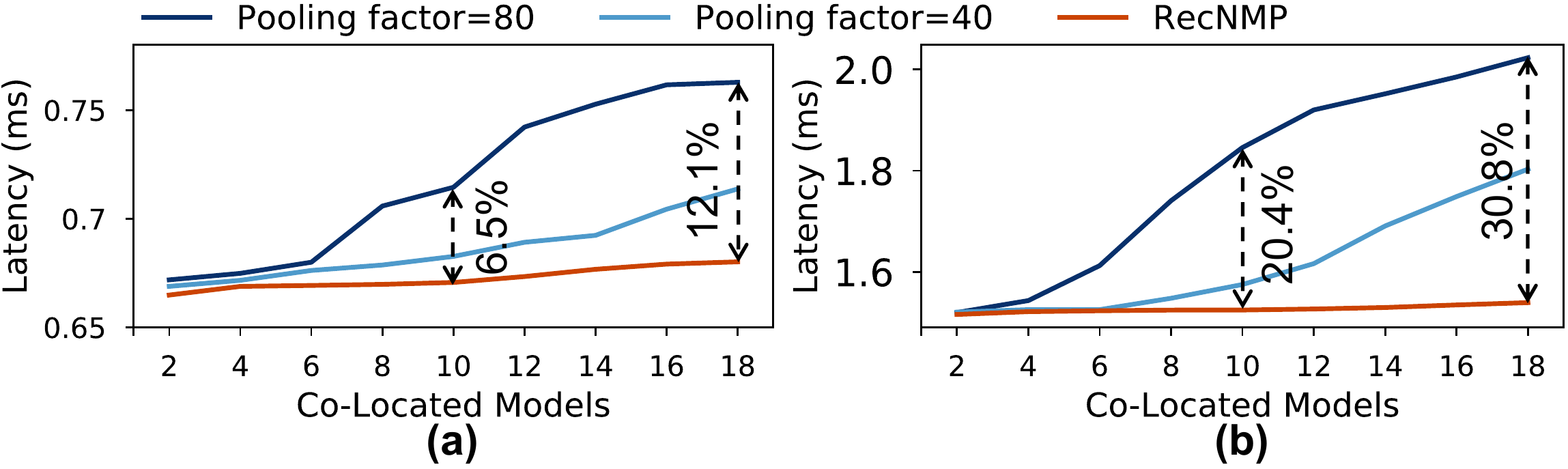}
  \vspace{-0.8cm}
  \caption{Effect of model co-location on latency of (a) TopFC in RM2-small model; (b) TopFC in RM2-large model. }
  \label{fig:fc_contention}
  \vspace{-0.3cm}
\end{figure}

Figure~\ref{fig:fc_contention} shows the degree of performance degradation on the co-located FC operations.
The amount of performance degradation experienced by the FC layers varies by the FC sizes, the degree of co-location, and the pooling values.
When examining the FC performance in baseline systems,
we observe worsening FC performance with larger FC weights at higher co-location degrees and higher pooling values.
\DesName effectively reduces the pressure from the cache contention, we show the base \DesName design but RecNMP-opt impacts FC performance equally as it offloads the same SLS computation. 
This beneficial effect ranging from 12\% to 30\% is more pronounced for larger FCs whose weight parameters exceed the capacity of the L2 cache and reside mainly inside the LLC cache.
For smaller FCs whose working set fits inside the L2 cache (e.g. all BottomFC and RM1's TopFC), the relative improvement is comparatively lower ($\sim4\%$).

\vspace{-0.2cm}
\subsection{End-to-end Model Speedup}
\begin{figure}[t!]
  \centering
  \includegraphics[width=\columnwidth]{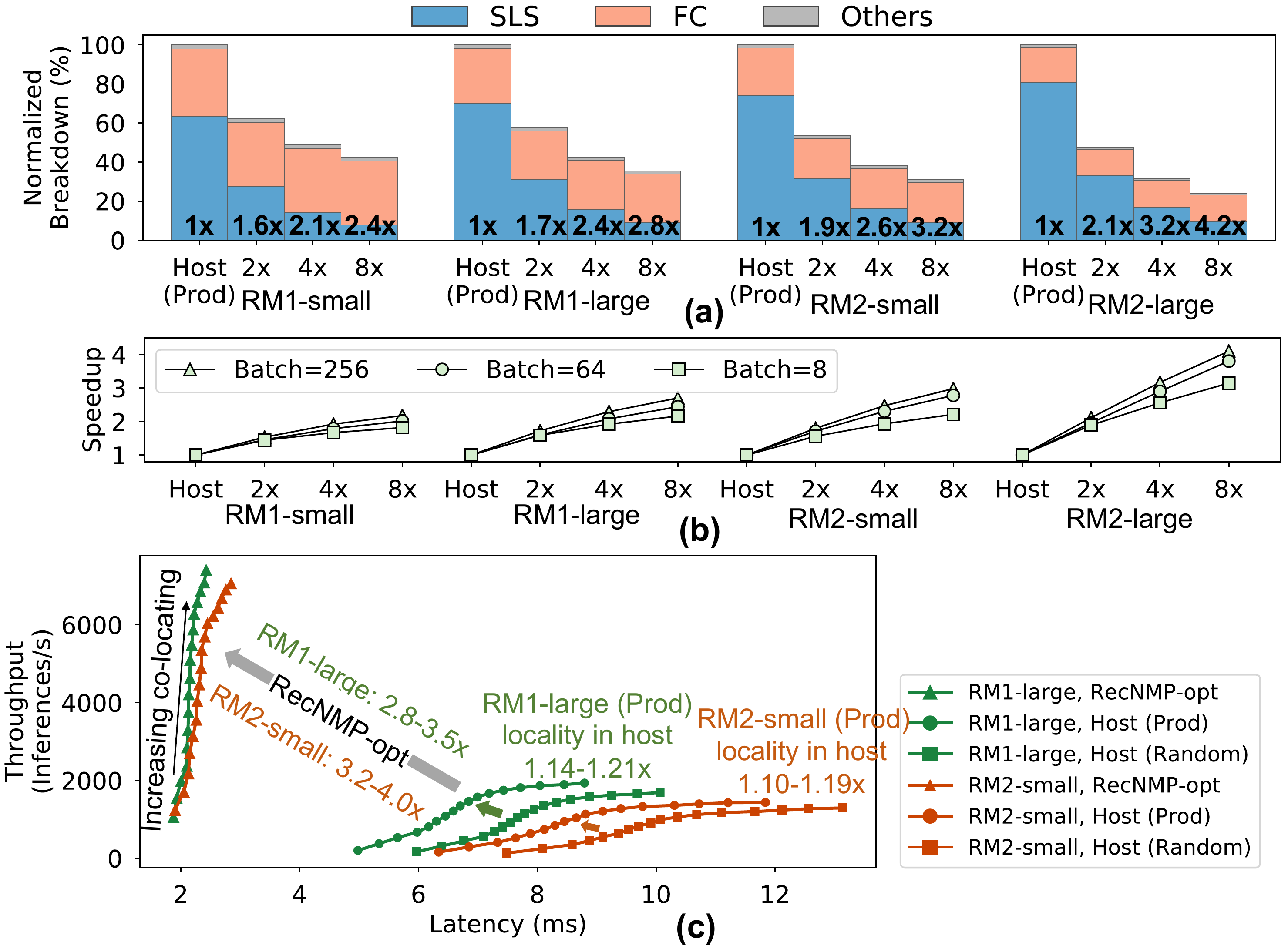}
  \vspace{-0.7cm}
  \caption{(a) Single end-to-end speedup of recommendation inference with 2-rank, 4-rank and 8-rank \DesName systems; (b) Single model speedup with different batch size; (c) Host and RecNMP-opt co-located model latency-throughput tradeoff.}
  \label{fig:model_speedup}
  \vspace{-0.6cm}
\end{figure}

\vspace{-0.1cm}
\textbf{Throughput improvement.} To estimate the improvement of end-to-end recommendation inference latency, we calculate the total speedup by weighting the speedup of both SLS and non-SLS operators. 
We measure model-level speedup across all four representative model configurations, shown in Figure~\ref{fig:model_speedup}(a).
Not surprisingly, the model that spends the most time running SLS operators (RM2-large) receives the highest speedup.
In Figure~\ref{fig:model_speedup}(b), the performance improvement obtained by \DesName varies with batch size.
In general, the model-level speedup increases with a larger batch size, as the proportion of time spent in accelerated SLS operators grows.

Figure~\ref{fig:model_speedup}(c) looks at the overall effect of increasing co-location in the presence of random or production traces for both the CPU baseline and our proposed \DesName solution.
Co-location generally increases the system throughput at the cost of degrading latency.
Compared to random traces, the locality present in production traces improves performance.
However, this locality performance ``bonus'' wears off as the level of model co-location increases due to the cache interference from the growing number of embedding tables in multiple models.
Applying \DesName in a 8-rank system results in 2.8-3.5$\times$ and 3.2-4.0$\times$ end-to-end speedup of RM1-large and RM2-small as the number of co-located models increases, because the fraction of SLS latency rises.
The improvement of both latency and throughput enabled by \DesName is clearly observed compared to the baseline system.

\textbf{Memory energy savings.} Comparing with the baseline DRAM system, \DesName provide 45.8\% memory energy saving.
\DesName saves the energy from the reduced data movement between the processor and the memory by performing local accumulation near DRAM devices and the leakage saving from reduced latency.
In addition, by incorporating memory-side caching and applying co-optimization techniques to improve RankCache hit rate, \DesName achieves extra energy savings by reducing the number of DRAM accesses.

\textbf{Area/power overhead.}
We estimate \DesName design overhead assuming 250MHz clock frequency and 40nm CMOS technology.
The area and power numbers are derived from Synopsys Design Compiler (DC) for the arithmetic and control logic and Cacti~\cite{cacti} for SRAM memory (i.e. RankCache).
Table~\ref{tab:design_overhead} summarizes the overhead of each \DesNameni processing unit for both the basic configuration without cache and the optimized configuration with cache optimization.

\begin{table}[]
\caption{Summary of \DesName Design Overhead}
\label{tab:design_overhead}
\centering

\begin{tabular}{|c|c|c|c|}
\hline
\multirow{2}{*}{} & \multicolumn{2}{c|}{RecNMP PU} & \multirow{2}{*}{\begin{tabular}[c]{@{}c@{}}Chameleon~\cite{chameleon}\\ (8 CGRA\\ accelerators)\end{tabular}} \\ \cline{2-3}
 & \begin{tabular}[c]{@{}c@{}}RecNMP-base\\ w/o RankCache\end{tabular} & \begin{tabular}[c]{@{}c@{}}RecNMP-opt\\ with RankCache\end{tabular} &  \\ \hline
Area (mm$^2$) & 0.34 & 0.54 & 8.34 \\ \hline
Power (mW) & 151.3 & 184.2 & 3138.6-3251.8 \\ \hline
\end{tabular}
\vspace{-0.5cm}
\end{table}

Compared with Chameleon, which embeds 8 CGRA cores per DIMM, our \DesNameni PU consumes a fraction of the area (4.1\%, 6.5\% for RecNMP-base and RecNMP-opt) and power (4.6-5.9\%).
When scaling \DesNameni PUs to multiple ranks in the DIMM, the total area and power will grow linearly, but it also translates to linearly-scaled embedding speedup.
Given that a single DIMM consumes 13W~\cite{tensorDIMM} and a typical buffer chip takes up 100mm$^2$~\cite{buffer-chip}, \DesName incurs small area/power overhead that can easily be accommodated without requiring any changes to the DRAM devices.

\vspace{-0.1cm}
\section{Related Work}
\vspace{-0.1cm}

\textbf{Performance characterization of recommendation models.}
Recent publications have discussed the importance and scale of personalized recommendation models in data center~\cite{DLRM, arxiv-gupta-19, mlperf, alibabaRec,sigarch-blog}.
Compared to CNNs, RNNs, and FCs~\cite{Minerva, EIE, Eyeriss,mlperf, fathom}, the analysis demonstrates how recommendation models have unique storage, memory bandwidth, and compute requirements.
For instance, ~\cite{arxiv-gupta-19} illustrates how Facebook's personalized recommendation models are dominated by embedding table operations.
To the best of our knowledge, \DesName is the first to perform locality study using production-scale models with representative embedding traces. 

\textbf{DRAM-based near-memory and near-data acceleration.}
Many prior works explore near-memory processing using 3D/2.5D-stacked DRAM technology (e.g. HMC/HBM)~\cite{nda-kim,graphpim,NeuroCube,3d1,3d2,3d3,3d4,3d5,3d6,3d7,3d8}. Due to their limited memory capacity ($16-32$GB) and high cost of ownership, these schemes are not suitable for large-scale deployment of recommendation models (10s to 100s of GBs) in production environment.
Chameleon~\cite{chameleon} introduces a practical approach to performing near-memory processing by integrating CGRA-type accelerators inside the data buffer devices in a commodity LRDIMM~\cite{chameleon}.
Unlike Chameleon's DIMM-level acceleration, \DesName exploits rank-level parallelism with higher speedup potential.
\DesName also employs a lightweight NMP design tailored to sparse embedding operators with much lower area and power overheads than CGRAs.

\textbf{System optimization for memory-constrained learning models.}
Sparse embedding representations have been commonly employed to augment deep neural network (DNN) models with external memory to memorize previous history. 
Eisenman et al. explore the use of NVMs for large embedding storage~\cite{eisenman2018bandana}. 
Although the proposed techniques result in $2-3\times$ improvement of effective NVM read bandwidth ($2.3GB/s$), it remains far below typical DRAM bandwidth ($76.8GB/s$) and cannot fundamentally address the memory bandwidth bottleneck in recommendation models.
MnnFast targets optimization for memory-augmented neural network and proposes a dedicated embedding cache to eliminate the cache contention between embedding and inference operations~\cite{mnnfast}.
However, these techniques do not directly apply to personalized recommendation consisting order-of-magnitude larger embedding tables.
TensorDIMM~\cite{tensorDIMM} proposes a custom DIMM module enhanced with near-memory processing cores for embedding and tensor operations in deep learning.
The address mapping scheme in TensorDIMM interleaves consecutive 64B within each embedding vector across the DIMM modules.
Its performance thus scales at the DIMM level and relies on the inherent high spatial locality of large embedding vectors, it is unable to apply to this approach to small vectors (e.g. 64B).
Given the same memory configuration, our design can outperform TensorDIMM in memory latency speedup by extracting additional performance gains from rank-level parallelism and memory-side caching optimizations.
The introduction of a customized compressed NMP instruction in \DesName also fundamentally addresses the C/A bandwidth constraints, without the restrictions on small embedding vectors as imposed by TensorDIMM.

\vspace{-0.2cm}

\section{Conclusion}
\vspace{-0.1cm}

We propose \textit{RecNMP}---a practical and scalable near-memory solution for personalized recommendation.
We perform a systematic characterization of production-relevant recommendation models and reveal its performance bottleneck.
A light-weight, commodity DRAM compliant design, \DesName maximally exploits rank-level parallelism and temporal locality of production embedding traces to achieve up to $9.8\times$ performance improvement of sparse embedding operation (carried out by the SLS-family operators).
Offloading SLS also offers alleviated cache contention for the non-SLS operators that remain in the CPU, resulting in up to 30\% latency reduction for co-located FC operators.
Overall, our system-level evaluation demonstrates that \DesName offers up to $4.2\times$ throughput improvement and 45.8\% memory energy saving with representative production-relevant model configurations.



\bibliographystyle{IEEEtranS}
\bibliography{ms}

\end{document}